\documentclass[prd,letterpaper,onecolumn,amsmath,amsfonts,amssymb,nofootinbib]{revtex4}
\pdfoutput=1
\usepackage {amsmath} 
\usepackage[dvips]{graphicx}        
\usepackage{feynmp} 
\usepackage{amssymb}          
\newcommand{\be}{\begin{equation}}
\newcommand{\ee}{\end{equation}}
\newcommand{\bear}{\begin{eqnarray}} 
\newcommand{\eear}{\end{eqnarray}}

\newcommand{\vev}[1]{\left\langle #1\right\rangle}

\newcommand{\lapproxeq}{\lower .7ex\hbox{$\;\stackrel{\textstyle
<}{\sim}\;$}}
\newcommand{\gapproxeq}{\lower .7ex\hbox{$\;\stackrel{\textstyle
>}{\sim}\;$}}
\newcommand{\stackdown}[2]{\lower 1.4ex\hbox{$\;\stackrel{\textstyle{#1}}
{\scriptstyle{#2}}\;$}}
\newcommand{\beq}{\begin{equation}}
\newcommand{\eeq}{\end{equation}}

\newcommand{\ba}{\begin{eqnarray}}
\newcommand{\ea}{\end{eqnarray}}

\newcommand{\bea}{\begin{eqnarray}}
\newcommand{\eea}{\end{eqnarray}}

\makeatletter
\def\slash{\@ifnextchar[{\fmsl@sh}{\fmsl@sh[0mu]}}
\def\fmsl@sh[#1]#2{%
  \mathchoice
    {\@fmsl@sh\displaystyle{#1}{#2}}%
    {\@fmsl@sh\textstyle{#1}{#2}}%
    {\@fmsl@sh\scriptstyle{#1}{#2}}%
    {\@fmsl@sh\scriptscriptstyle{#1}{#2}}}
\def\@fmsl@sh#1#2#3{\m@th\ooalign{$\hfil#1\mkern#2/\hfil$\crcr$#1#3$}}
\makeatother
\usepackage{color}

\definecolor{orange}{rgb}{0.9,0.2,0}
\definecolor{brown}{rgb}{0.7,0.3,0.2}
\definecolor{fuxia}{rgb}{1,0,1}
\definecolor{skyblue}{rgb}{0,0.1,0.9}
\definecolor{violetred}{rgb}{0.8,0.13,0.56}
\definecolor{deeppink}{rgb}{1.00,0.08,0.5}
\definecolor{pink}{rgb}{1.00,0.75,0.80}
\definecolor{orchid}{rgb}{0.85,0.44,0.84}
\definecolor{lightpink}{rgb}{1.00,0.71,0.76}
\definecolor{bluish}{rgb}{0,0.6,0.8}  
\begin{document}

\title{Deforming the Starobinsky model in ghost-free higher derivative supergravities  

\vspace*{8mm}
{\it Dedicated to the memory of Peggy  Kouroumalou, colleague and friend } 
}

\author{
\vspace*{8mm}
{\bf G. A.~\ Diamandis} {\footnote{email: gdiam@phys.uoa.gr}}, \, {\bf B. C.~\ Georgalas}{\footnote{email: vgeorgal@phys.uoa.gr}}, \, {\bf  K.~\ Kaskavelis}{\footnote{email: kkaskavelis@phys.uoa.gr}}, \, 
 \\ {\bf A. B.~\ Lahanas} {\footnote{email: alahanas@phys.uoa.gr}} and \, {\bf  G.~\ Pavlopoulos}{\footnote{email: gepavlo@phys.uoa.gr}}}
\affiliation{National and Kapodistrian University of Athens, Department of Physics,\\
Nuclear and Particle Physics Section, GR--157 71  Athens, Greece}

\vspace*{2cm}   
\begin{abstract}
We consider higher derivative supergravities that are dual  to ghost-free $N=1$ supergravity theories in the Einstein frame. 
The duality is implemented   by deforming the K\"ahler function, and/or the superpotential,  to include nonlinear dependences on  
chiral fields that in other approaches play the role of the Lagrange multipliers employed to establish this duality. 
These models are of the  no-scale type, and in the minimal case,  require the presence of four chiral multiplets, 
with a  K\"ahler potential having the structure  of the  $ SU(4,1)/SU(4) \times U(1) $ coset manifold.   In the standard  $N=1$ supergravity formulation, these models are described by a multifield scalar potential, featuring Starobinsky-like behavior in particular  directions.

\end{abstract}
\maketitle


\section{Introduction}
The study of generalizations of Einstein gravity,  considering higher order curvature terms, has a long standing history, for various reasons, related  to either cosmology, or towards  the effort for understanding the ultraviolet behavior of gravity at the quantum level. The recent activity on this field is mainly motivated by Starobinsky's model of inflation \cite{STAR} and, in particular, by the fact that the inflaton may  have a ``dual" description as the extra scalar mode propagating in a $R+R^2$ theory \cite{stelle,whitt}. Strictly speaking it is proven that this theory is equivalent to Einstein gravity specifically coupled to a scalar field.  
Going beyond the $ \sim R^2$,  general $F(R)$-theories have been studied, see for instance  \cite{CAPO,SOTIRIOU} and references therein,  which are known to be equivalent to the Einstein-Hilbert action,  if one introduces additional auxiliary fields which couple to curvature in the  Jordan frame. By appropriate Weyl rescalings the Jordan theory   is brought to the Einstein-Hilbert Lagrangian  with the auxiliary fields becoming dynamical. 
Also some classes of gravity theories, whose Lagrangians are not only  functions of the curvature but may include 
$ \Box^n R$ terms, had been considered in the past.  These higher derivative gravity theories have been proven to be equivalent to the Einstein-Hilbert action with additional  scalar fields \cite{wands}. 
In another context, higher derivative gravities  have been invoked  against improving the UV (Ultraviolet) behavior of gravity theories, but  they suffer, in general, by the presence of negative norm states.  
This  issue has been analyzed in literature, where  general gravity actions were considered, involving terms at most quadratic in the Riemmann tensor, in an effort to obtain better UV-behavior and avoid having negative norm states \cite{GHOST,BISWAS1, BISWAS2}.  Further attempts towards constructing finite and  ghost-free, nonlocal \cite{MOD1} and local \cite{MOD2} gravity theories have been also considered. 

Minimal $N=1$ supergravity theories, \cite{CREMMER, BAGGER}, being the supersymmetric completion of the Einstein theory,  have  also been extensively studied in the last forty years or so,  in an effort to encompass all  of the known forces of Nature, including gravity, into a unified framework.        
On the other hand the supersymmetric completion of $R+R^2$ gravity,  pertinent to cosmological inflation,  is a  notable paradigm of how a dual description of the supersymmetric Einstein-Hilbert action can be accomplished. 
Driven by these, there is a strong motivation towards studying  supersymmetric completion of general $F(R)$-theories, going beyond the simple  $R+R^2$ model. 
The organization of Lagrangians involving  higher powers   and derivatives of the scalar curvature has been  addressed in the past.  The minimal $R + R^2$ theories were considered in \cite{THEISEN,CECOTTI}, which were shown to be  equivalent to standard supergravity coupled to two chiral supermultiplets.  Besides,  in \cite{CECOTTI} a general methodology was developed  in order to include arbitrary powers of the scalar curvature as well, which is accomplished using a set of chiral fields that play the role of Lagrange multipliers. In these approaches the dual supersymmetric $F(R)$-theories  are equivalent to  standard supergravity theories with  K\"{a}hler potentials of the no-scale type \cite{NOSCA, EKN}. 
  The main problem in adopting the full equivalence between the two descriptions, that is the higher-$R$ and the standard supergravity,  in the case the former  departs from the minimal $R^2$ theory, is that while in the higher-$R$ description there are no propagating ghost states at the linearised level. This is not the case in the dual description due to the specific form of the K\"{a}hler  function employed to implement the duality. 
  
  Recently there has been an intense activity towards building models that encompass  cosmological inflation,  especially after the precise data on the cosmological parameters  delivered by Planck and other collaborations \cite{PLANCK,BICEP,PLANCK2,BICEPPLANCK}.  The physics of inflation will be placed under further scrutiny,  in the next round of measurements,   and this led many authors to consider various models,  with or without supersymmetry, \cite{KALLOSHX,KETOV1,KETOV,ELLISp1,KALLOSHp1,SFERRARA2,KALLOSH,SFERRARA,ELLIS,KEHAGIASp1,KEHAGIAS2,DALIAN,SCALISI,ZWIR,PALLIS,KOUNNAS,KTAL,BUCH,ANTODUDAS,ovrut1,DGKKLP,SAGNIOTTI,teradax,YAMADA,ASAK,KOSHE}.  In this context supergravity and higher derivative supergravities may play a central role. 
   
  In this work we consider chiral higher derivative supergravities that  are dual  to ghost-free $N=1$ supergravities. This is done by deforming the Lagrangian  to include nonlinear dependences, on some of the would be Lagrange multipliers used in the approach described in  \cite{CECOTTI},  which in this way become dynamical.  Interestingly enough, even in the simple cases considered, this approach leads, in turn, to scalar potentials which in particular directions have a Starobinsky-like form.    
  
  This paper is parametrized as follows. In the following section we review the general set-up for the formulation of supergravity chiral actions,  which is an essential  tool towards building $F(R)$-supergravity theories,  and establish their duality to the standard $N=1$ supergravity description. This approach uses   a set of chiral fields, that appear linearly in the action, and play the role of Lagrange multipliers.  In Sec. III we apply this formalism focusing on theories that involve at most  two derivatives of the curvature $R$,  and discuss the problem associated with the ghost issue. In Sec. IV we proceed to a particular deformation of the model by promoting one of the Lagrange multipliers to a dynamical field, which therefore  is not eliminated any longer from the action. This is necessary in order to avoid ghosts in the ordinary description of the  $N=1$ supergravity.  Deforming the theory in this way adds extra difficulty,  in expressing the theory in chiral form, and the way this is implemented is discussed in detail.  Particular models are presented  in Sec. V, whose K\"ahler potential is reminiscent of the no-scale type. 
    In Sec. VI we analyze the mass spectrum of these models, and  in Sec. VII we consider their corresponding  $N=1$ supergravity description, in the Einstein frame,  and show that the scalars,  in a suitable superfield basis, destabilize a 
  $ SU(4,1)/SU(4) \times U(1) $ coset space. Their mass  spectrum  has no ghosts and exactly matches  that of the dual theory,  derived in the previous section.  Moreover we show that the scalar potential of the standard $N=1$ supergravity Lagrangian is positive definite with  a Minkowski vacuum with unbroken supersymmetry. This potential is described by four complex scalar fields 
and  in  particular directions is reminiscent of the  well-known Starobinsky model. It is for this reason that we have dubbed the class of models considered in this work as deformed Starobinsky models, although we are aware that the virtues of the single - field  Starobinsky model, as far as  cosmological inflation is concerned, are hard to obtain. 
Models of this type can only lead to behavior inflation, and the presence of additional scalars may stabilize the Starobinsky inflationary trajectory. Recently \cite{addazi}, extensions of the $R + R^2$ theory, in the framework of the $N=1$ old-minimal supergravity, were considered which are ghost-free, with one scalar field present and a stable potential.  
  A detailed analysis on the cosmological consequences of the class of models discussed in this work is not pursued here, and it will be presented in a forthcoming publication.

\section{Chiral Lagrangians}
In the absence of gauge fields and using superfield formalism, the $N=1$ Supergravity (SUGRA)  Lagrangian is written as 
\bea
{\cal{L}} \, = \, \int \, \, d^4 \Theta \, E^{-1} \, \Omega(S, \bar{S} ) \, + \, 
\left( \; 
\int \,  d^2 \Theta \, 
2 \, {\cal{E}} \,  
W(S) 
+ (H.c.) \; \right) 
\label{sug1}
\eea
In this   $S$ denotes collectively all chiral multiplets involved, which are  coupled to gravity,  and   $  \bar{S}  $  their  corresponding anti-chiral multiplets. The kinetic function $\Omega$  is  real  and   the superpotential $W$  is  a holomorphic  function . 
This Lagrangian, expressed in terms of components, is a function  of a K\"{a}hler function, given below, and  its derivatives, 
\bea
{\cal{G} } \, = \, {\cal{K} } \, + \, \ln \,  { | {{W}} |  }^{2}\,.\, 
\label{kaler}
\eea
In this  the  K\"{a}hler function  $  {\cal{K} }$ is related to $\Omega$ by  
\bea
{\cal{K}} \, \equiv \, -3 \, \ln \, \left(  - \dfrac{ \Omega }{ 3} \right) \,.\, 
\label{KKK}
\eea

The above  Lagrangian can be cast in the so called chiral form 
\bea
{\cal{L}} \, = \, \int \, \, d^2 \Theta \, 
2 \, {\cal{E}} \, \left(  \; 
- \dfrac{1}{8} \, ( \overline{\cal{D}  }^2 - 8 \, {\cal{R}} \, ) \,  \Omega(S, \bar{S} )  +{{ W}}(S)  
 \right) 
+ (H.c.) \; 
\label{sug2}
\eea
This is particularly  useful since  any $N=1$ supergravity action  can be written as a chiral action which  includes chiral multiplets and their corresponding  kinetic multiplets  !  We shall make extensive use of this form when dealing with higher $R$ supergravities.

In a previous paper \cite{DLT} we considered $F(R)$minimizavities whose construction is implemented using chiral multiplets coupled to gravity. The gravity sector itself is well-known to be described by  the chiral superspace density  $ {\cal{E}} $, 
the  supervierbein determinant  $E$,  and  the  chiral superspace curvature ${\cal{R}}  $ . 
Ignoring their fermionic components, $ {\cal{E}}  $ and $ {\cal{R}}  $    are given by 
\bea
{\cal{E}} \, & = & \,  \dfrac{e}{2} \, \left( 1 - \Theta^2 \,  \overline{M} \right) \nonumber \\
{\cal{R}} \, & = &  \,  - \dfrac{M}{6}  + \Theta^2 \, \left(  \dfrac{R}{12}  - \dfrac{ M \, \overline{M} }{9}  - \dfrac{b_\mu^2}{18}  
+ \dfrac{i }{6} \, D_\mu b^\mu \,   \right)  ,
\label{mults}
\eea
where $M$ and $b_\mu$ are the auxiliary fields of the gravity multiplet.
{\footnote{
Throughout this paper, the metric signature is  $\, - + + +  $ and the sign of the scalar  curvature $\, R$
 coincides with that used in  \cite{CREMMER}. 
}}
In superfield formalism, omitting fermions, a Poincar\'e chiral multiplet is written as  
\bea
\Phi \, = \, A + \Theta^2 \; F .
\eea
Given a chiral multiplet  $\, \Phi $, another  chiral multiplet can be constructed whose scalar component includes $\overline{F}$, that is the complex conjugate of $F$.  This is called kinetic multiplet and is given by 
{\footnote{
This is $\, - 1 / 4$ times the  corresponding multiplet used in reference \cite{BAGGER}. 
}}
\bea
 T(\Phi) \, &=& \, ( \,  \overline{F \,} - \dfrac{M}{3} \,    \overline{A \,} \, ) 
 \nonumber \\
&+ & 
 \Theta^2 \left(  \Box \,  \overline{A \,} +   \dfrac{i}{3} \, D_\mu b^\mu \,  \overline{A \,}  +   \dfrac{2 \,i}{3} \,  b^\mu \, \partial_\mu  \overline{A \,}   - \dfrac{b_\mu^2}{9} \,  \overline{A \,} + \dfrac{R}{6} \,  \overline{A \,} + \dfrac{2  \overline{M \,}}{3 } \,( \,  \overline{F \,} - \dfrac{M}{3} \,  \overline{A \,}  )    \right) .
\eea
In this  $\Box$ is the ordinary gravity d' Alembertian operator, i.e.  $  \Box = \frac{1}{e}  \,    \partial_\mu ( e g^{\mu \nu} \partial_\nu )$.
This multiplet can be expressed as the chiral  projection of the anti-chiral field $  \overline{\Phi}$ , 
\bea
- \, \dfrac{1}{4} \; ( \, {\overline{\cal{D}}}^{\, 2} \, - \, 8 \, {\cal{R}} \, ) \, \overline{\Phi} \, = \, T(\Phi) .
\label{proj}
\eea
With this definition the kinetic multiplet of the unit chiral multiplet $\, \Sigma_0 = 1$ is twice the curvature multiplet, i.e. 
\bea
T(\Sigma_0) \, = \, \left( \, - \dfrac{M}{3} \,  \right) +
 \Theta^2 \left(  \,  \dfrac{R}{6}  +  \dfrac{i}{3} \, D_\mu b^\mu \,    
 - \dfrac{b_\mu^2}{9} \,  \,  - \dfrac{2 | M |^2}{9 } \,
   \right) 
   \, = \, 2 \, {\cal{R}} 
\eea
while the kinetic multiplet of the curvature chiral multiplet  is given by
\bea
T( {\cal{R}}) \, &=& \, 
\left( \,  \dfrac{R}{12} -  \dfrac{b_\mu^2}{18} - \dfrac{i}{6} \, D_\mu b^\mu \, - \dfrac{{ | M |}^2}{18 \, } \, \right) \, + \, 
\Theta^2 \, 
\left(  \, 
\dfrac{\overline{M}}{36} \, R - \dfrac{{\Box \, \overline{M} } }{6} \, - \dfrac{{ | M |}^2 }{27} \,  \overline{M}
\, \right.
\nonumber \\
  &  & \quad - \left. \dfrac{\overline{M} }{  54} \, b_\mu^2 - \dfrac{i \, \overline{M} }{6} \,  \,D_\mu b^\mu
   - \dfrac{i}{9} \,  b^\mu \, \partial_\mu \overline{M}  
  \right)
\eea
Later, we shall use  the kinetic multiplet of  $ T( {\cal{R}}) $ which, in a straightforward manner, is found to be 
\bea
T(T( {\cal{R}})) \, &=& \, 
\left(  - \, \dfrac { \Box \, M}{ 6 } \, - \,  \dfrac {  \, M | M |^2}{ 54 }    + \dfrac{i}{9} \, M \, D_\mu b^\mu \, + \, 
 \dfrac{i}{9} \,  b^\mu \, \partial_\mu M
 \right) +
\Theta^2 \, 
\left(  \, \dfrac{1}{72} \, \left( R - \dfrac{2}{3} b_\mu^2 \, \right)^2 \, + \, \dfrac{1}{12} \, \Box \,\left( R - \dfrac{2}{3} b_\mu^2 \, \right)
\, \right.
\nonumber \\
  &  & 
\quad \, + \, \dfrac{1}{18} \, \left(  - \dfrac{ | M |^2}{6} + i \, D_\mu b^\mu \right) \, \left( R - \dfrac{2}{3} b_\mu^2 \, \right) +
 \dfrac{\; b^\mu}{18} \, \partial_\mu \, \left( R - \dfrac{2}{3} b_\mu^2 \, \right)
 \, - \, \dfrac{\Box \, | M |^2}{18 } \, - \, \dfrac{\overline{M} \Box \,  M }{9 } \, - \, \dfrac{ | M |^4}{ 81  } 
 \nonumber \\
 & &
\quad  \left.
 \, + \, \dfrac{i}{18} \, | M |^2 \, D_\mu b^\mu \, + \, \dfrac{i}{27} \, b^\mu \, ( \, 2 \, \overline{M} \,  \partial_\mu M - \partial_\mu \, | M |^2 \,  ) \, + \, \dfrac{i}{6} \, \Box \, D_\mu b^\mu \, - \, \dfrac{1}{9} \, b^\nu \, \partial_\nu \,  D_\mu b^\mu \, - \, 
 \dfrac{1}{18} \, ( D_\mu b^\mu)^2
  \right) \hspace*{0.7cm}
\eea
Note that this includes $\, R^2$ and $\, \Box\, R$ within its $\, F$-term.  These forms are the important  building blocks  towards building  $\, F(R)$-supergravity theories, as we shall see. 


One can build actions involving the chiral multiplets $ {\cal{R}}   $  ,  $ T({\cal{R}}  )  $,  $ T(T({\cal{R}}  ))  $, and so on, as well as other chiral multiplets,  which we denote collectively by  $X$.  Thus one may  consider  Lagrangians  having  the general form, 
\bea
{\cal{L}}_0 \, = && \int \, d^4 \Theta \, E^{-1} \, 
\Omega_0( X , \bar{X},  {\cal{R}} , \overline{ {\cal{R}} } ,  T({\cal{R}}) , \overline{ T({\cal{R}})} , \cdots) 
\nonumber \\
&& +
\left( \, 
\, \int \,  d^2 \Theta \,  \, 2 {\cal{E}} \,  W_0( X ,{\cal{R}}, T({\cal{R}}), T(T({\cal{R}})), ...) \,  + \, (H.c.)  .
\right) .
\label{chi3}
\eea
Such Lagrangians describe  higher $R$ theories,  by construction,  and these are equivalent to  standard $N=1$ supergravities, in which only the Einstein term appears \cite{CECOTTI}.  In fact  by introducing Lagrange multipliers,  $\,  \Lambda, \Lambda_1, \Lambda_2 \, \cdots $, 
one can show that  (\ref{chi3}) is equivalent to  a standard   $N=1$ supergravity  described by the following functions,
\bea
\Omega \, &=& \, \Omega_0 ( X , \bar{X} , J_1 , \overline{J_1} , J_2 , \overline{J}_2 , \cdots ) 
\nonumber \\
&& - \left( \, \Lambda +  \, \overline{\Lambda} \,  \right) 
 \, - \, 2 \left( \, \Lambda_1  {\overline{ J_1}   } \,  +  \, J_1 \, {\overline{\Lambda}}_1 \,  \right) \, \cdots \, 
 \, - \, 2 \left( \, \Lambda_{n}  {\overline{ J}_{n}  } \,  +  \, J_{n} \, {\overline{\Lambda}}_{n} \,  \right) 
 \nonumber \\
W \, &=& \, 
  W_0( X, J_1, J_2,  ...  ) \, + \, 2 \, \Lambda \, J_1 \, + \, 2 \, \Lambda_1 \, J_2 \, \cdots \,  + 2 \, \Lambda_n \,  J_{n+1} .
  \label{o22}
\eea
In these neither $\Omega_0$ nor $W_0$ depend on $\, \Lambda, \Lambda_1, \Lambda_2 \, \cdots  $ .  
The theory described by  $\Omega, W \, $ is then written as
\bea
{\cal{L}}  = {\cal{L}}_0  + {\cal{L}}_{\Lambda} ,
\label{ssppll}
\eea
where   $ {\cal{L}}_{\Lambda}  $ is  the  part of the Lagrangian dependent on the  Lagrange multipliers,  appearing in the function 
$\Omega$ and the superpotential $\, W$ given before in Eq. (\ref{o22}),   
\bea
{\cal{L}}_{\Lambda} \, &=& \, 2 \, \int \,  d^2 \Theta \,  \, 2 {\cal{E}} \,
\left[ \;    \Lambda \, ( \, J_1 - {\cal{R}} \, ) 
\, + \,   \Lambda_1 \, ( \, J_2 - T(J_1)  \, ) \, 
\,  \right.
\nonumber \\
&& \quad \quad 
\left.
\, + \,   \Lambda_2 \, ( \, J_3 - T(J_2)  \, ) \, +   \cdots 
\, + \,   \Lambda_n \, ( \, J_{n+1} - T(J_{n})  \, )
\,  + \cdots \right] \, + \, (H.c.) ,
 \label{multipl}
\eea
In deriving this, repeated use was made of the very important relation 
\bea
&& \int \, \, d^4 \Theta \, E^{-1} \, ( \, S \overline{H} + \overline{S} \, H \, ) \, = \,   
\nonumber \\
 && \hspace*{1.5cm}  \dfrac{1}{2} \; \int \,  d^2 \Theta \,  2 \, \, {\cal{E}} \,  \; ( S T(H) + H T(S) ) \, + (H.c.)  \, = \, 
\int \,  d^2 \Theta \, \,  2 \,  {\cal{E}} \,  \; (\,  S \,T(H) \,) \, + (H.c.) ,
\label{ident}  
\eea
which  holds true, up to  four divergences, for any  chiral multiplets $\, S, H $. Its derivation is almost straightforward, using  the equivalence of the Lagrangians (\ref{sug1}) and  (\ref{sug2}) and employing  (\ref{proj}).  
Solving with respect the Lagrange multipliers $ \Lambda , \Lambda_1 , \Lambda_2 \cdots \Lambda_{n} $ we get
\bea
J_1 = {\cal{R}} \, , \, J_2 = T(J_1)=T({\cal{R}}) \, , \, J_3 = T(J_2)=T(T({\cal{R}})) \, \quad  \cdots \quad J_{n+1}=T(J_n) = T(...T({\cal{R}})...) .
\label{sol3}
\eea
Plugging  the solutions $\, J_1, J_2 ...  $    into (\ref{ssppll}) yields exactly  (\ref{chi3}), due to the fact that 
 $ {\cal{L}}_{\Lambda}  $ vanishes.   This proves the equivalence of the two theories. 

 As an instructive well-known example consider the no-scale supergravity \cite{CECOTTI,KETOV,KALLOSHp1,ELLISp1,SFERRARA2} model described by
\bea
\Omega = -3 \,  ( \, T + \overline{T} - \Phi \overline{\Phi} \,  ) 
\quad , \quad W = 3 \, \mu \, \Phi \left( T - \dfrac{1}{2} \right) .
\label{nscale}
\eea
One sees that in this case we have one Lagrange multiplier $\Lambda$, which equals to   $3 T$ in this case, while $J_1$ is identified with 
$\, \mu \, \Phi / 2$. The functions $\Omega_0, W_0$ are given by
\bea
\Omega_0 = \, 3 \, \Phi \overline{\Phi} \quad , \quad W_0 = - \dfrac{ 3 \,\mu }{2} \, \Phi .
\eea
The Lagrangian (\ref{multipl}) is, in this case, given by  
\bea
{\cal{L}}_{\Lambda} \, = \, \int \,  d^2 \Theta \,  \, 2 {\cal{E}} \,
\left[ \;  W_0( \Phi ) + \, 6 \, T \, ( \, \dfrac{\mu}{2} \, \Phi - {\cal{R}} \, ) \right] \, + \, (H.c.) ,
\eea
and the total Lagrangian, cast in chiral form, is
\bea
{\cal{L}} \, = \, \int \,  d^2 \Theta \,  \, 2 {\cal{E}} \,
\left[ \;    \dfrac{3}{2} \, \Phi \, T(\Phi) \, + \,
W_0( \Phi ) + \, 6 \, T \, ( \, \dfrac{\mu}{2} \, \Phi - {\cal{R}} \, ) \right] \, + \, (H.c.) .
\label{chi55}
\eea
The equation of motion for  $T$,  $ \,  {\delta {\cal{L}} } / {\delta T} =0 \, $,  is  
\bea
\Phi = \dfrac{2}{\mu} \, {\cal{R}} ,
\label{solv1}
\eea
which when  plugged into  the Lagrangian (\ref{chi55})  eliminates the last term, which is proportional to $T$, leaving  
\bea
{\cal{L}} \, = \, \int \,  d^2 \Theta \,  \, 2 {\cal{E}} \,
\left[ \;    \dfrac{6}{\mu^2} \, {\cal{R}}  \, T({\cal{R}} ) \, - \, 3 \,  {\cal{R}}   \right] \, + \, (H.c.) .
\label{chi66}
\eea
Using the explicit  forms of $ \, {\cal{E}} ,  {\cal{R}} ,   T(  {\cal{R}}   )  \,    $,  given previously,  this  trivially  leads to the following Lagrangian, also derived in  \cite{DLT} using the component formalism, 
\bea
 e^{ \, -1} \, {\cal{L}}_{dual}  \, & = &\, \,   - \, \dfrac{R}{2} \, + \, \dfrac{R^2  }{  12 \, \mu^2 } \, 
 - \,    \dfrac{1}{9 \mu^2}  \, \left(  \, \dfrac{| M |^2 }{2} +  {b_\mu^2 } \, \right) \, R \,  
 - \,  \dfrac{1}{3 \mu^2}   \,  {|  \nabla_\mu \, M \, |}^2
 \nonumber \\
 && \, +\,   \dfrac{\, | M |^4 }{27 \mu^2} \, -  \dfrac{\, | M |^2}{3 } 
 -  \,  \dfrac{i }{9 \mu^2}\, b_\mu \, ( \overline{M} \,  \nabla_\mu \, M - c.c \, ) \nonumber \\
 &&\, + \, \dfrac{1}{3 \, \mu^2} \, { ( D_\mu \, b^\mu  ) }^2  \, +   \, \dfrac{b_\mu^4 }{27 \, \mu^2} 
 \, + \,   \dfrac{ b_\mu ^2}{3 } \,   + \,  \dfrac{ b_\mu ^2}{27 \mu^2 }  \, | M |^2  .
 \label{staror}  
\eea
Note that the  scalar $M$ is linearly  related to the scalar component $ \phi$ of the superfield $\Phi$, on account of (\ref{solv1}). In fact 
$\, M = - 3 \mu \, \phi$.
The term$\sim R^2$ arises from the first term of (\ref{chi66}) and the linear term  $- R / 2$ from the second term of the same Lagrangian. 
\footnote{
Modulo  stabilization terms,  introduced to stabilize the inflationary trajectory, Eq. (\ref{staror})  is the one obtained in \cite{addazi} when $M / 6$ is replaced by $X$, used in that work,  and the constant  $f_0$  of that paper is taken vanishing.
}
 
This is the dual form of the ordinary  $N=1$  supergravity Lagrangian specified by $\Omega, W$ given in (\ref{nscale}) .
The bosonic part  of the latter depends on $T, \Phi$ scalars and is linear in the curvature $R$, describing   therefore six degrees of freedom (d.o.f.), the same as (\ref{staror}).  Along the  direction $\Phi = 0$ this receives a  simple form 
\bea{
e^{ \, -1} \, {\cal{L}} \, = \, 
- \, \dfrac{R}{2} \, - \dfrac{ 3 \, { | \nabla_\mu \, T  |  }^2 }{   {( T + \bar{T}  )}^2  } \, 
- \, 3 \, \mu^2 \,   \dfrac{  \, { | T - 1/2 |  }^2 }{   {( T + \bar{T}  )}^2  } 
}\,.
\eea
If  $Im T$ is frozen  to  $Im \, T = 0   $,  then by defining a canonically normalized field $\varphi$, by
$  \, Re \, T \, \equiv  \, \frac{1}{2} \, e^{ \sqrt{ \frac{2}{3}  }  \, \varphi }   $,  we get 
\bea
e^{ \, -1} \, {\cal{L}} \, = \, 
- \, \dfrac{R}{2} \, - \dfrac{1}{2} \, { ( \nabla_\mu \, \varphi   ) }^2  \, 
- \, \dfrac{ 3 \, \mu^2 }{ 4  } \, { \left( \, 1 - e^ { - \, \sqrt{2/3} \, \varphi  }  \right)   }^{\,2} ,
\eea
which is the celebrated Starobinsky's model \cite{STAR,whitt} , with $\varphi$ being the inflaton field with the parameter  $\mu$ setting the scale of inflation. 

\section{Higher derivative theories}
The strategy outlined in the previous section can be employed to construct higher derivative supergravity theories. The supersymmetric Starobinsky model includes terms at most quadratic in the curvature  $R$.  Using   two Lagrange multipliers we can  build a higher derivative $F(R)$-theory,  as follows. Consider  the theory described by
\bea
\Omega &=&  \, T + \overline{T} + ( Q \overline{\Phi} \, + \, \Phi \overline{Q}) + 
\omega(X, \overline{X}, \Phi, \overline{\Phi},C, \overline{C})
\nonumber \\
W &=& T \Phi + Q C + h(X, \Phi, C) .
\label{nscale2}
\eea
The correspondence with the previous notation, if needed, is given by
\bea
T = - \Lambda, \; \Phi = - 2 J_1, \; Q = \Lambda_1, \; C = 2 J_2 .
\eea
The functions $\omega, h$ do not involve, at this stage, any of  the Lagrange multipliers $T, Q$. Treating the Lagrange multipliers as described  in the previous section,  the Lagrangian corresponding to this theory has the following form, 
\bea
{\cal{L}} \, &=& \, \int \, \, d^4 \Theta \, E^{-1} \, \omega(X, \overline{X}, \Phi, \overline{\Phi},C, \overline{C}) \, 
\nonumber \\
&& 
 \; 
+ \, 
\left( \, \int \,  d^2 \Theta \,  2 \, {\cal{E}} \,  
 \left( \; h(X, \Phi, C) \, + \, T \, ( \, \Phi + 2  {\cal{R}} \, ) \, + \, Q \, ( \, C + T(\Phi) \,) \;
 \right)
+ (H.c.) \; \right) .
\label{yxxx1}
\eea
This is easily solved for the Lagrange multipliers  $T, Q$, 
\bea
\Phi = - 2 {\cal{R}} \quad , \quad  C = - T(\Phi) =  2 \, T({\cal{R}}) .
\eea
With $\Phi , C   $  plugged into (\ref{yxxx1}) we get 
\bea
{\cal{L}} \, &=& \, \int \, \, d^4 \Theta \, E^{-1} \, \omega(X, \overline{X}, - 2 {\cal{R}} , - 2 \overline{{\cal{R}}}, 2 \, T(  {\cal{R}}) , 
2 \,  \overline{T(  {\cal{R}}) }) \, 
\nonumber \\
&&
 \; 
+ \, \int \,  d^2 \Theta \,  2 \, {\cal{E}} \,  
  \; h(X, - 2 {\cal{R}} ,   2 \, T(  {\cal{R}})) \, 
+ (H.c.) \; .
\label{yxxx2}
\eea
We  remark that using only two Lagrange multipliers is sufficient to build higher-$R$ supergravities. This is the most economic way 
and in the rest of this work  we shall not pursue more complicated cases involving a larger number of them. 

In order to implement this,  consider,  as an example, the simple case of having a theory specified by 
\bea
\omega = 2 \alpha \, C \, \overline{C} \quad , \quad h = 0 .
\label{ccc} 
\eea
According to (\ref{yxxx2}) this theory is 
\bea
{\cal{L}} \, &=& \, 8 \alpha \,  \int \, \, d^4 \Theta \, E^{-1} \, T(  {\cal{R}}) \,   \overline{T(  {\cal{R}}) } \, 
= \,
4 \alpha \; 
\int \,  d^2 \Theta \,  2 \, \, {\cal{E}} \,  \;
 T(  {\cal{R}}) \, T(T(  {\cal{R}}))
+ (H.c.) \; ,
\label{TTR}
\eea
where in the last step we put the Lagrangian in chiral form as prescribed in the previous section. 
This is  a higher-$R$ theory.  In order to show this, collect the terms that depend only on the curvature,  after integrating over 
$\Theta^2$. The result is 
\bea
e^{-1} \, {\cal{L}} \, = \, \dfrac{\alpha}{18} \, \left( \, \dfrac{R^3}{6} \, + \, R \, \Box R \,  \right) \, + \, \text{other terms } ,
\label{RRR}
\eea
which is indeed a higher $R$ theory.  

However it should be noted that in the ordinary  $N=1$ supergravity form,  the theory described by  (\ref{nscale2}) is linear in
 $\, \sim R  $, and  ghosts exist that are not present  in the dual description (\ref{TTR}). 
 The appearance of  ghost states becomes  manifest by  the fact that some of the eigenvalues of the complex scalar kinetic matrix are negative. On the other hand counting the degrees of freedom of the two theories there is a mismatch. The dual theory appears with fewer degrees of freedom as compared to the ordinary  $N=1$ supergravity and no ghosts at all.  
  Obviously the ghost states of $N=1$ supergravity  should decouple,  in some manner,  since the number of physical degrees of freedom in both descriptions should be equal. 
 A  way to implement the ghost decoupling,  in some particular cases, was given in \cite{DLT}.  Here we shall pursue an alternative way by constructing theories that have no ghosts  in their  $N=1$ formulation. This is merely  done by sacrificing the role of the field $Q$ as being a Lagrange multiplier.   In doing that,  the field  $Q$ becomes dynamical and is no longer eliminated, and then  the degrees of freedom in the two descriptions match. In the following section we shall give the details of how this is implemented and chiral actions of this kind can be constructed.

\section{$Q$- Deformations}

A way to circumvent the ghost problem, outlined in the closing remarks of the previous section, is to deform the theory so that the functions $\, \omega$ 
and or $\, h$ in  Eq. (\ref{nscale2}) depend on the Lagrange multiplier $\, Q$. As we already remarked this  may rectify the situation and no ghost states appear in the standard formulation of the  $N=1$ supergravity. 

In order to tackle the problem in the most general  manner  let us consider theories in which the functions $\Omega, W$ are given by
\bea
\Omega &=&  \, T + \overline{T} + ( Q \overline{\Phi} \, + \, \Phi \overline{Q}) + 
\omega(X, \overline{X}, \Phi, \overline{\Phi}, C, \overline{C},  Q, \overline{Q} )
\nonumber \\
W &=& T \Phi + Q C + h(X, \Phi, C, Q) .
\label{nscale3}
\eea
These are like (\ref{nscale2}), however, the functions $\omega, h$ are now allowed to have $Q$ dependencies.  Evidently the  dependence on the superfield  $Q$ is not linear any longer,  and hence, $Q$ is not eliminated from the action.  

We shall assume the most general form for the function $\, \omega$ which when expanded in powers of the chiral  superfields 
$\,  X,  \Phi,  C, Q $, and their associated  anti-chiral fields $\,  \overline{X}, \overline{\Phi},  \overline{C}, \overline{Q} $, it receives the form
\bea
\omega \, = \, \sum \, a_{\, n_i m_j  }  \, ( \, X^{n_1} \Phi^{n_2} C^{\, n_3} Q^{n_4} \,) \, 
( \, \overline{X}^{\, m_1} \overline{\Phi}^{\, m_2} \overline{C}^{\, m_3} \overline{Q}^{\, m_4} \,)
\, + \, (H.c.) .
\eea
 Using (\ref{ident})  this can be written as 
\bea
\int \, \, d^4 \Theta \, E^{-1} \, \omega \, = \, 
 \; \int \,  d^2 \Theta \,  2 \,  {\cal{E}} \,  \sum \, f(X,\Phi, C, Q) \,  T(g(X,\Phi, C, Q) )
 \, + \, (H.c.) , 
 \label{zzz1}
\eea
where the sum extends over the monomials 
\bea
f  \, \equiv \, a_{\, n_i m_j  }  \, \, X^{n_1} \Phi^{n_2} C^{\, n_3} Q^{n_4} \,
\quad , \quad 
g  \, \equiv \,   \, \overline{X}^{\, m_1} \overline{\Phi}^{\, m_2} \overline{C}^{\, m_3} \overline{Q}^{\, m_4} 
\eea
With that done the Lagrangian can be put in chiral form 
\bea
{\cal{L}} \, &=& \int \,  d^2 \Theta \,  \,  2 \, {\cal{E}} \,  \, P( X, \Phi, C, Q  ) \, + \, (H.c.) ,
\label{chir2}
\eea
with the superpotential function $\, P$ given by 
\bea
P \, &=& \, h(X, \Phi, C, Q) \, + \, T \, ( \, \Phi + 2  {\cal{R}} \, ) \, + \, Q \, ( \, C + T(\Phi) \,)  
\nonumber \\
&& + \,\sum \, f(X,\Phi, C, Q) \,  T(g(X,\Phi, C, Q) ) .
\label{xxx4}
\eea
Denoting for convenience
\bea
H \, \equiv h(X, \Phi, C, Q) \, + \, T \, ( \, \Phi + 2  {\cal{R}} \, ) \, + \, Q \, ( \, C + T(\Phi) \,) , 
\label{HHH}
\eea
any superfield variation  $\delta A$ of the action, where $A$ is any of $X, \Phi, C, Q, T $,  yields
\bea
\delta \,  \int \,  d^2 \Theta \,  \,  2 \, {\cal{E}} \,  \, P \,& =&  \,
  \int \,  d^2 \Theta \,  \,  2 \, {\cal{E}} \, \left(   
 \dfrac{\partial H }{ \partial A} \, \delta A \, + \,
 \sum \left[ \; f( A + \delta A , ...) \, T( g( A + \delta A , ...  )) \, - \,  f(A, ...)  T(g(A ...)) \, \; \right]
 \right)
 \nonumber \\
&=&  \int \,  d^2 \Theta \,  \,  2 \, {\cal{E}} \, \left(   
 \dfrac{\partial H }{ \partial A} \, \delta A \, + \,
 \sum \dfrac{\partial f }{ \partial A} \,  \, T( g ) \, \delta A  
 + \sum  f \, T\left( \dfrac{\partial g }{ \partial A} \,  \delta A \right) 
 \right)
  \nonumber \\
&=&  \int \,  d^2 \Theta \,  \,  2 \, {\cal{E}} \, \left(   
 \dfrac{\partial H }{ \partial A}  \, + \,
 \sum 
 \left( \, \dfrac{\partial f }{ \partial A} \,  \, T( g ) \,  
 + \dfrac{\partial g }{ \partial A} \,  \, T( f ) \, \right)
 \right)  \, \delta A  .
\eea
In the last step we used the fact that 
\bea
 \int \,  d^2 \Theta \,  \,  2 \, {\cal{E}}  A T(B) \, = \,  \int \,  d^2 \Theta \,  \,  2 \, {\cal{E}}  B T(A) \, ,
\eea
which holds true  up to four divergences. Therefore the equations of motion for $A$, in superfield form, are given by 
\bea
\dfrac{\partial H }{ \partial A}  \, + \,
 \sum 
 \left( \; \dfrac{\partial f }{ \partial A} \,  \, T( g ) \,  
 + \dfrac{\partial g }{ \partial A} \,  \, T( f ) \, \right)  = \, 0 .
 \label{eom1x}
\eea

In order to proceed further let us consider  a specific  function $\omega$ given by 
\bea
\omega = 2\, \alpha \, C \, \overline{C} + 2 \, \lambda \, Q \overline{Q} + 
2 \, \beta \, \Phi \,  \overline{\Phi} ,
\label{om1}
\eea
where the constants $\, \alpha, \beta , \lambda$ are assumed positive. The function $\omega$ in Eq. (\ref{om1}) ensures that the  standard supergravity $N=1$  theory has a positive kinetic function in the sector $C, \Phi, Q$, under the condition  
$4 \beta \lambda - 1 > 0$, and therefore no ghosts appear ! 
For the case at hand the  following terms are encountered in the sum $\, \sum \, f \, T(g)  $ of Eq. (\ref{zzz1}) with the $f, g$ terms  given, respectively,  by 
\bea
f &=& \alpha \, C \quad , \quad g = C
\nonumber \\
f &=& \lambda \, Q \quad , \quad g = Q
\nonumber \\
f &=& \beta \,  \Phi  \, , \quad \quad g = \Phi .
\eea
In the following we assume, for simplicity, that there is no $Q$ dependence of the superpotential function $h$. 
Also, we assume that no additional multiplets $X$  are present. These can be added in a trivial manner later, if desired. 
Then  the equations of motion for the superfields $T, Q$, using equation (\ref{eom1x}) and the form of $\, H$ given by  Eq. (\ref{HHH}), yield
\bea
\dfrac{\partial }{\partial T} :  \quad \quad && \Phi + 2 \, {\cal{R}} =0 
\nonumber \\
\dfrac{\partial }{\partial Q} :  \quad \quad && C + T(\Phi) + 2 \lambda T(Q)  \, = 0 .
\label{eoms1}
\eea
Solving for $\, \Phi, C$ and  plugging  into the function $\, P$, given in Eq. (\ref{xxx4}), one arrives at  
\bea
P \, &=& \, h( \Phi, C) \, + \alpha \, C \, T(C) - \, \lambda \, Q \, T(Q) \, + \, \beta \, \Phi \, T(\Phi)  .
\label{xxx6}
\eea
In this $\, \Phi, C$ are solved by (\ref{eoms1}),  that is they are given by 
\bea
\Phi =  - \, 2 \, {\cal{R}}  \quad , \quad C = 2 \, ( T( {\cal{R}}  )  - \lambda T(Q)) .
\label{sol2}
\eea
We see that the field $Q$, unlike $T$,  is no longer eliminated and was not expected to. 
The independent chiral multiplets are $\,  {\cal{R}}  \, , \, Q   $ and the Lagrangian is expressed in terms of these,  and  kinetic multiplets that follow from these chiral superfields.  Actually,  from Eqs. (\ref{chir2}) and (\ref{xxx4}) we find that the resulting chiral Lagrangian  has the form, 
\bea
{\cal{L}} \, &=& \int \,  d^2 \Theta \,  \,  2 \, {\cal{E}} \,  \, P( \Phi, C, Q  ) \, + \, (H.c.)
\nonumber \\
&=& \int \,  d^2 \Theta \,  \,  2 \, {\cal{E}} \, \left[ \, h(\Phi, C) + 4 \alpha \, T( {\cal{R}} ) \, T(T( {\cal{R}} ))  
 + 4 \beta \,  {\cal{R}}   \, T( {\cal{R}} ) - 8 \alpha \lambda \, T( {\cal{R}} )  \, T(T( Q ))  \right.
\nonumber  \\
 &&
 \left.
 + \, 4 \, \alpha \lambda^2 \, T( Q)  \, T(T( Q )) - \lambda \, Q T(Q)
  \right] \,  + \, (H.c.) ,
\label{chir22}
\eea
where it is meant that the arguments within  $h(\Phi, C)$ are replaced by the solutions given in  (\ref{sol2}). 
This is certainly a higher $F(R)$-supergravity. In fact  we have seen from  (\ref{TTR}) and  (\ref{RRR}),  that the term 
$ \, \sim \, T( {\cal{R}} ) \, T(T( {\cal{R}} ))   $ leads to 
\bea
e^{-1} \, {\cal{L}} \, = \, \dfrac{\alpha}{18} \, \left( \, \dfrac{R^3}{6} \, + \, R \, \Box R \,  \right) \, + \, \cdots
\label{high1}
\eea
where the ellipsis denote additional  terms that either mix $R$ with other fields or they do not depend on the curvature at all. 
Note the appearance of the term $\, R \Box R$ which is unavoidable due to the appearance of the $\, C \, \overline{C}$ term in the definition of the function $\, \omega$, see Eq. (\ref{om1}).  Note also the appearance of the term 
$\,  \sim \,   {\cal{R}}   \, T( {\cal{R}} )  $,  encountered also  in the Starobinsky action (\ref{chi66}), which  yields, see  (\ref{staror}), 
\bea
e^{-1} \, {\cal{L}} \, = \, \dfrac{\beta}{18} \,  \, R^2 \,  + \, \cdots
\label{high2}
\eea
However this does not exhaust all possibilities and the presence   of the superpotential function $\, h$  in Eq. (\ref{chir22})  is source of additional $R$-dependent terms yielding higher-$R$ supergravities. In the following section,  we shall consider particular choices for the function $h(\Phi, C)$, some of which are generalizations of the Starobinsky model.

\section{Building $F(R)$-Supergravities}

In this section we shall consider specific models,  whose the pertinent functions $\, \Omega, W$  are as given in (\ref{nscale3}), with the function $\omega$ defined  by (\ref{om1}).  The function $h$ assumed  to depend  only on  $\Phi, C$, that is the superpotential part $h$ has no $Q$-dependence.   As we shall see the deviation from the $Q$-linearity, existing in the function $\omega$,   induces  deformations of the Starobinsky model for  properly chosen functions  $\, h(\Phi, C)$. 

From Eqs. (\ref{sol2}) we have for the scalar components $\, \phi, c$ and the corresponding  $F$-terms,  $\, F_\phi, F_c$ of the chiral fields $\, \Phi, C$ 
\bea
c \, &=& \, \dfrac{R}{6} - \dfrac{b_\mu^2}{9} - \dfrac{i}{3} \, D_\mu b^\mu - \dfrac{ | M |^2 }{ 9 } - 2
 \, \lambda \, \left( \,   \overline{F_q} - \dfrac{M}{3} \, \overline{q} \right)  
 \label{solccc1}
 \\
F_c &=& \left( \,    \dfrac{ \overline{M} }{ 18 } \, R  - \dfrac{ \Box \,   \overline{M}   }{ 3 }  - \dfrac{ 2 \,  \overline{M} \, | M |^2  }{  27 }    
- \dfrac{ \overline{M}  }{ 27  }   \, b_\mu^2 \, - \, i \, \dfrac{ \overline{M} }{3} \, D_\mu b^\mu - \,  \dfrac{ 2 \, i }{ 9  } \, b^\mu \partial_\mu \overline{M} \,  \right) 
 \nonumber  \\
&- & 2 \lambda \, \left( \left( \Box + \dfrac{R}{6} \right) \overline{q} + \dfrac{i}{3} \, D_\mu b^\mu \, \overline{q}  - \dfrac{b_\mu^2}{9}  \, \overline{q} + \dfrac{ 2 \, i }{ 3  } \, b^\mu \partial_\mu  \overline{q} +  \dfrac{ 2 \, \overline{M} }{ 3 } \,  \left( \,  \overline{F_q} - \dfrac{M}{3} \, \overline{q} \right) 
\right) ,
\label{solccc2}
\eea
In these $\, q, F_q$ stand for the scalar component and the $F$-term of the chiral multiplet $\, Q$.  
The solutions for the components of the multiplet $\Phi$ is much easier to handle  since $\Phi$ is  just  $\, - 2 \,{\cal{R}} $,  yielding 
\bea
\phi = \dfrac{ M}{3  } 
\quad , \quad
F_\phi = - \, \dfrac{R}{6} + \dfrac{b_\mu^2}{9} - \dfrac{i}{3} \, D_\mu b^\mu + \dfrac{ 2 \, | M |^2 }{ 9 }  \, .
\label{solfff1}
\eea
The Lagrangian (\ref{chir22}) involves terms that are products of two multiplets and thus we can make use of the general result
\bea
{\cal{L}} =
 \int \,  d^2 \Theta \,  \,  2 \, {\cal{E}} \;  A \,  B \, + (H.c.) \, = \,   
 e \, \left( \,    - \overline{M} \, a b + ( \, a \,F_b + b \, F_a \, )      \,  \right) + (H.c.) .
 \label{pro1}
\eea
This is   easily derived, ignoring the fermionic contributions. $\, A , B$ are any two  chiral multiplets, whose scalar components  are 
 denoted by the lower case letters $a, b$,  while their  $F$-terms are denoted by  $\, F_a, F_b $  respectively. 
 
 \subsection{Models with  $h = f(C)$}

Let us first consider a simple model in which the superpotential part involves a general function of the multiplet $C$, that is
\bea
h \, = \, f(C) .
\eea
In this case, writing  the multiplet $C$  as $\, C = c + \Theta^2 \, F_c  $,  the function    $h$ receives the form
\bea
h(C) \, = \, f(c) \, + \, \Theta^2 \, f^\prime(c) \, F_c .  
\label{super1}
\eea
Then using  (\ref{pro1}), taking one of the multiplets to be the unit multiplet,  we easily  get for the $h$-dependent part of the Lagrangian, 
\bea
{\cal{L}}_h  \equiv 
 \int \,  d^2 \Theta \,  \,  2 \, {\cal{E}} \;   \,  h(C) \, + (H.c.) \, = \,   
 e \, \left( \,    - \overline{M} \, f(c) +  \, f^\prime(c) \, F_c     \,  \right) + (H.c.) .
 \label{onlyc}
 \eea
Using (\ref{solccc1}) and (\ref{solccc2}), and collecting the terms in $\,  f(c), \, f^\prime(c)$ and $F_c$ 
that depend only on the curvature $R$, we get, 
\bea
e^{\, -1} \, {\cal{L}}_h  \, = \,   
\overline{M} \, \left(   - f(R/6) \, +  \,  \dfrac{R}{18} \, f^\prime(R/6)  \,   \right) 
\, - \, \lambda \, \dfrac{R}{3} \,  f^\prime(R/6) \,  \overline{q} + (H.c.) .
\label{elin}
\eea
In this $\, q$ is the scalar component of $\, Q$.  Evidently this  does not contain pure curvature dependent terms. It involves terms in which the curvature mixes with other fields, namely $\, M , q$ in this case. 
The same holds for the remaining terms of $\, e^{\, -1} \, {\cal{L}}_h  $ that we have not shown.  As we shall see,  in the Lagrangian    (\ref{chir22}) all fields are dynamical, even $F_q$. 
Therefore $M, q$ cannot be expressed in terms of other fields and  (\ref{elin}) cannot lead to a Lagrangian  depending exclusively on the curvature $R$.  Therefore this choice for the superpotential $h$ leads to a dual supergravity description whose pure  $R$-terms are only those presented  in (\ref{high1}), (\ref{high2}).  Obviously one needs to depart from this type of superpotential $h$ in order to build dual supergravities involving higher powers of the curvature, other than  those given  in (\ref{high1}), (\ref{high2}). 

\subsection{Models with  $h = \Phi \, f(C)$}

From the previous discussion we have seen that the function $h$ should involve, besides the dependence on $C$, dependence on the superfield $\Phi$ as well, in order to construct a general $F(R)$-supergravity. An interesting case arises when   $h$  is linear in the superfield $\Phi$ having the form 
\bea
h(\Phi, C) \, = \, \Phi \, f(C) ,
\label{hfi}
\eea
Interestingly enough this choice leads to generalizations of the supersymmetric Starobinsky model.  We shall term these as ``deformed" Starobinsky models.  In order to see this, and to make contact with the usual  notation found  in literature, we  rescale the fields  $\Phi$ and $T$ fields by 
\bea
T \, \rightarrow \, - \, 3 \, T \quad , \quad \Phi \, \rightarrow \, - \, \mu \, \Phi
\quad , \quad \text{where} \quad \mu \equiv \sqrt{ \dfrac{3}{2 \, \beta}  } \; .
\label{resss}
\eea
Then the functions $\Omega , W$ given in (\ref{nscale3}), with $\omega$ defined by (\ref{om1})  and the function  $h$ given by   (\ref{hfi}),  take the following forms 
\bea
\Omega &=&  \, - 3 \, ( T + \overline{T} - \overline{\Phi} \, \Phi ) 
- \, \mu \,( \, Q \overline{\Phi} \, + \, \Phi \overline{Q} \, ) + 
2  \lambda  \, \overline{Q} \, Q \, + \, 2 \alpha \,  \overline{C} \, C 
\nonumber \\
W &=& 3 \, \mu \, \Phi \, \left( \, T - \dfrac{1}{2} \, \right)  + Q C - \mu \, \Phi \, \Sigma(C) ,
\label{defstar}
\eea
where $\,  \Sigma(C)  \equiv f(C) - 3/2$.  
The function $\Omega$ gives rise to a K\"ahler potential having the structure of the no-scale models. The symmetries of the associated K\"ahlerian manifold will be discussed later. 
The first terms of $\Omega, W$ above, depending on $\Phi, T$, are the ones encountered in (\ref{nscale}), the supersymmetric Starobinsky model. 
As we shall see later,  the standard $N=1$ supergravity description of  this model has a  Starobinsky-like potential along a particular direction. However the scale of the scalar potential of the inflaton field  is not $ \mu^2 $, although it is related to it. We shall come to this point later.

The class of models just  discussed are higher $R$-supergravities in their dual description.  
The presence of   $Q$ and $C$ kinetic terms in $\omega$, which are necessary  in order to ensure absence of ghost states in the standard   $\, N = 1$ supergravity description,   induces  terms  higher than   $\sim R + R^2$  encountered in the  simple Starobinsky model.  
Actually, we have already seen that   $\, \sim R^3 + R \, \Box \,  R $ terms are induced, see Eq.( \ref{high1}),  due to the appearance of  
the $  \overline{C} \, C $ term in $\omega$. However the presence of a nontrivial superpotential part $h$, as given above,  gives rise to additional curvature dependent terms leading to more general $F(R)$-supergravities. 

In order to find the curvature dependent terms,  stemming from $\, h$, we shall  consider $f(C)$ in (\ref{hfi}) to be an arbitrary function of the chiral field $C$.  In this case the chiral form of  the superpotential $h$ is 
\bea
h(\Phi, C) \, = \, \phi \, f(c) \, + \, \Theta^2 \, ( \,  f(c) \, F_\phi + \phi f^\prime (c)  \, F_c \, ) , 
\eea
so that, in this case,  we get  from the $h$-dependent part of the Lagrangian given in (\ref{chir22})
\bea
{\cal{L}}_h  = 
 \int \,  d^2 \Theta \,  \,  2 \, {\cal{E}} \;   \,  \Phi \, h(C) \, + (H.c.) \, = \,   
 e \, \left( \,    - \overline{M} \, \phi \, f(c) +  \,  F_\phi \, f(c)  + \, \phi \, f^\prime(c) \, F_c     \,  \right) + (H.c.) ,
 \label{onlyc2}
 \eea
Replacing  in this the solutions (\ref{solccc1}) to (\ref{solfff1}) we get, in a straightforward manner, the $h$-dependent   part of the Lagrangian which is given by, 
\bea
e^{\, -1} \, {\cal{L}}_h  &=&
\left( \, - \,  \dfrac{R}{6} + \dfrac{b_\mu^2}{9} - \dfrac{i}{3} \, D_\mu b^\mu - \dfrac{ | M |^2 }{ 9 }   \right)  \, f(c) 
\nonumber \\
&\;  +& \, f^\prime(c) \; \left( 
 \left( \,    \dfrac{ \; | M |^2 }{ 54 } \, R  - \dfrac{  M \, \Box \,   \overline{M}   }{ 9 }  - \dfrac{ 2 \,  | M |^4  }{  81 }    
- \dfrac{  | M |^2 }{ 81  }   \, b_\mu^2 \, - \, i \, \dfrac{ | M |^2 }{9} \, D_\mu b^\mu - \,  \dfrac{ 2 \, i }{ 27  } \, b^\mu 
\, M \partial_\mu  \overline{M} \,  \right) \right.
\nonumber \\
&\; -&
\left. 
  \lambda \, \dfrac{ 2 M}{3} \,  
 \left( 
 \left( \Box + \dfrac{R}{6} \, \right)  \overline{q} + \dfrac{i}{3} \, D_\mu b^\mu \,  \overline{q}  - \dfrac{b_\mu^2}{9}  \,  \overline{q}
+ \dfrac{ 2 \, i }{ 3  } \, b^\mu \partial_\mu   \overline{q} +  \dfrac{ 2 \, \overline{M} }{ 3 }   \left( \,  \overline{F_q} - \dfrac{M}{3} \,  \overline{q} \right) 
 \right)
\right) 
\nonumber \\
\, &+& \, (H.c.) .
 \label{fffccc}
 \eea
In this it is meant, without saying, that the  scalar field $c$, appearing within both $\, f(c) , f^\prime(c)$,   is expressed in terms of other fields using  the solution  (\ref{solccc1}).  Putting $\lambda = 0$ we get the result derived  in \cite{DLT}, see Eq. (45) in this reference. 
However $\, \lambda \neq 0$ is mandatory in order to have an Einstein supergravity without ghosts, as we have already remarked.
If we keep the terms that depend only on the curvature $R$, in the above Lagrangian, and adding the corresponding curvature dependent contributions  from (\ref{high1}), (\ref{high2})  we arrive at 
\bea
e^{\, -1} \, {\cal{L}} \, = \, - \, \dfrac{R}{3} \, f \left( \dfrac{R}{6}  \right ) \, + \,
\dfrac{\alpha}{18} \, \left( \, \dfrac{R^3}{6} \, + \, R \, \Box R \,  \right) \,  
+ \dfrac{\beta}{18} \,  \, R^2 \, + \, \cdots
\label{rrrttt}
\eea
In this, for convenience, we have taken $f(c)$ to be real function when its argument $c$ is real. The ellipsis denotes additional terms, among them  curvature dependent  terms which however  mix with other fields. 
The Lagrangian (\ref{rrrttt})   is indeed a $F(R)$-supergravity whose precise form is  specified by the choice of the function  $f(c)$. 
For the simple  choice $\, f(C) = 3/2$, which eliminates the last term in the superpotential given in Eq.  (\ref{defstar}),   the deformed Starobinsky model leads to 
\bea
e^{\, -1} \, {\cal{L}} \, = \, - \, \dfrac{R}{2} \, + \, \dfrac{R^2  }{  12 \, \mu^2 } \, +
\dfrac{\alpha}{18} \, \left( \, \dfrac{R^3}{6} \, + \, R \, \Box R \,  \right) \,
 + \, \cdots
 \quad \text{with} \quad   \mu = \sqrt{ \dfrac{3}{2 \, \beta}  }  .
\eea
For the complete form of the Lagrangian one should also add to  (\ref{rrrttt}) the terms from Eq.  (\ref{chir22}) that depend on the chiral multiplet $Q$. These do not contribute to terms that depend solely on the curvature $R$. However they are essential for studying the mass spectrum of the dual theory. This task will be undertaken in the following section.

\section{The mass spectrum of the dual $F(R)$-theory }

The mass spectrum of the dual theory can be read by isolating the bilinear terms in the Lagrangian (\ref{chir22}). 
To that purpose, we shall pick the quadratic in the fields  terms, separately for  each term appearing within  (\ref{chir22}),  keeping however the complete expressions for  those  terms that depend exclusively on the curvature $R$. 

Using previous results, see Eqs.   (\ref{chi66}) and (\ref{staror}),  from the term  
$\, \sim {\cal{R}}   \, T( {\cal{R}} )  $ we get 
\bea
&& {\cal{L}}_1  \equiv  
\int \,  d^2 \Theta \,  \,  2 \, {\cal{E}} \,  {\cal{R}}   \, T( {\cal{R}} )  + (H.c.) \, = \, 
\nonumber \\
 &  & \hspace*{2.5cm} \, 
 e \, \left( \, \dfrac{R^2  }{  72 } \, 
 - \,    \dfrac{1}{54}  \, \left(  \, \dfrac{| M |^2 }{2} +  {b_\mu^2 } \, \right) \, R \,  
 - \,  \dfrac{1}{18}   \,  {|  \nabla_\mu \, M \, |}^2  \, + \,   \dfrac{\, | M |^4 }{162} \right.
 \nonumber \\
& &  \hspace*{2.5cm} 
\left.
\, - \,   \dfrac{i }{54}\, b_\mu \, ( \overline{M} \,  \nabla_\mu \, M - c.c \, ) 
 \, + \, \dfrac{1}{ 18} \, { ( D_\mu \, b^\mu  ) }^2  \, +   \, \dfrac{b_\mu^4 }{ 162 } 
 \, \,   + \,  \dfrac{ b_\mu ^2}{ 162 }  \, | M |^2  \, \right) .
 \label{staror2}
\eea
This is the complete expression.  Collecting the quadratic terms, with the exception of the terms that are only $R$-dependent, as we have already remarked, we get
\bea
e^{\, -1} \, {\cal{L}}_1^{\, (quad)}   = 
 \, \dfrac{R^2  }{  72 } \, 
 - \,  \dfrac{1}{18}   \,  {|  \nabla_\mu \, M \, |}^2  \
 \, + \, \dfrac{1}{ 18} \, { ( D_\mu \, b^\mu  ) }^2  \,  .
 \label{quad1}  
 \eea
For the term in the Lagrangian (\ref{chir22}) having the structure
\bea
 {\cal{L}}_2  \equiv  
\int \,  d^2 \Theta \,  \,  2 \, {\cal{E}} \,  T({\cal{R}} )  \, T(T( {\cal{R}} ) )+  (H.c.) ,
\eea
 the quadratic pieces are given by
 \bea
 e^{\, -1} \, {\cal{L}}_2^{\, (quad)}   = 
 \dfrac{1}{72} \, \left( \,   \dfrac{R^3}{6} + R \,  \Box \, R     \, \right) 
 + \dfrac{1}{18 } \,  ( D_\mu \, b^\mu  )  \, \Box \, ( D_\mu \, b^\mu  ) 
 \, + \,  \dfrac{1}{18 }  \,  \Box \, M \; \Box \, \overline{M} .
  \label{quad2} 
 \eea
 For the  term which mixes $  T({\cal{R}} )  $  with $ T(T( Q ) ) $, namely  
\bea
 {\cal{L}}_3  \equiv  
\int \,  d^2 \Theta \,  \,  2 \, {\cal{E}} \,  T({\cal{R}} )  \, T(T( Q ) ) +  (H.c.) ,
\eea
 the quadratic terms are given by
 \bea
 e^{\, -1} \, {\cal{L}}_3^{\, (quad)}   = 
 \dfrac{R}{12 } \, \, \Box \, ( F_q +  \overline{F_q} ) \, - \, \dfrac{i}{6} \,  D_\mu \, b^\mu  \,   \Box \, ( F_q -  \overline{F_q}  ) 
 \, - \, \dfrac{1}{6} \, ( \, \Box \,  \overline{q} \, \Box \, M + H.c.  \, ) .
  \label{quad3}
 \eea
 As for the terms that depend on the multiplet $Q$, the term
\bea
 {\cal{L}}_4  \equiv  
\int \,  d^2 \Theta \,  \,  2 \, {\cal{E}} \,  Q  \, T(Q ) +  (H.c.) 
\eea
 yields a quadratic piece given by 
 \bea
 e^{\, -1} \, {\cal{L}}_4^{\, (quad)}   = q \, \Box \,  \overline{q} +  \overline{q} \, \Box \, q \, + \, 2 \,  \overline{F_q} \, F_q ,
  \label{quad4}
 \eea
while  the term   
\bea
 {\cal{L}}_5  \equiv  
\int \,  d^2 \Theta \,  \,  2 \, {\cal{E}} \,  T(Q)  \, T(T(Q ))  +  (H.c.) 
\eea
gives rise to quadratic terms given by
 \bea
 e^{\, -1} \, {\cal{L}}_5^{\, (quad)}   = 
 \overline{F_q} \, \Box \, F_q \, + \, F_q \, \Box \, \overline{F_q} \, + \,  2 \,  \Box \, \overline{q}  \, \Box\, q .
  \label{quad5}
 \eea
 
 Note that the  fields $\, F_q$ in the dual description, unlike the Einstein frame supergravity,  are no longer auxiliary and hence cannot  be eliminated ! This is intimately related to the fact that nonlinear $\, Q \overline{Q}$ terms were introduced. Therefore, the deformed theory has additional dynamical  d.o.f., as compared to a theory in which $\, Q$ appears linearly, and thus  it comes closer to having the same number of d.o.f.  with the ordinary $\, N=1$ supergravity in the Einstein frame. In fact,  there is no mismatch in the number of d.o.f., as we shall see.,  which is a welcome feature signaling the absence of ghosts in the standard 
   $\, N=1$ supergravity description. 
 
 It only remains to read the quadratic part of the Lagrangian (\ref{fffccc}) which can be implemented by expanding the function $f(c)$. 
 By a straightforward calculation one finds,  recalling that  the function  $f(c)$ has been taken  real for real values of  $c$, 
 \bea
e^{\, -1} \, {\cal{L}}_h^{(quad)}   &=&
- \, \dfrac{R}{3} \, f \left( \dfrac{R}{6} \right) \, + \, \dfrac{2}{9} \,   \left(\, {b_\mu^2}  - { | M |^2 }  \right) \, f(0) 
\nonumber \\
&&
\left( \,   - \dfrac{2}{9} \,   ( D_\mu \, b^\mu  )^2  - \dfrac{1}{9} \, ( \,  M \, \Box \, \overline{M} + \overline{M}  \, \Box \, M \, ) 
+ \lambda \, \dfrac{R}{3} \, ( F_q + \overline{F_q} ) - \dfrac{2 \, i \lambda}{3} \, D_\mu b^\mu \, ( F_q - \overline{F_q} ) \right.
\nonumber \\
&& \left.  \hspace*{3mm} - \, \dfrac{2  \lambda}{3} \, ( \,  M \, \Box \, \overline{q}+ \overline{M} \, \Box \, q \, ) 
 \,  \right) \, f^\prime(0) .
 \label{quadh}
 \eea

In the last step we should collect all quadratic terms, given so far, in order to read the mass spectrum of this dual theory. This may not be as easy due to field mixings occurring in the Lagrangian.   In doing so it proves easier to use the real and imaginary components of the fields involved as follows
\bea
F_q = S + i G \;   , \;  M / 3 = A + i B \;   , \;  q = \rho + i \sigma
\quad \text{and also} \quad 
D_\mu b^\mu = \Psi .
\label{newfield}
\eea
The longitudinal component of the field $\, b^\mu$ we have denoted by $\, \Psi$. With these definitions the quadratic part of the total Lagrangian  is given by 
 \bea
 e^{\, -1} \, {\cal{L}}^{\, (quad)}   &= &  F(R) \, + 
  \dfrac{\alpha}{18} \,  R \,  \Box \, R  \,  
  - \, \dfrac{4 \, \alpha \lambda}{3} \, R \, \Box \, S \, + \,   \dfrac{2  \, \lambda \, f_0^\prime}{3} \,  \, R  \, S
  \nonumber \\
& + &  \dfrac{2 \, \alpha }{9} \, \Psi \, \Box \, \Psi \, + \,  \dfrac{2 \, ( \beta - f_0^\prime )     }{9} \, \Psi^2 \, - \, 
\dfrac{8 \, \alpha \lambda}{3} \, \Psi \, \Box \, G \, + \, \dfrac{4 \,  \lambda \, f_0^\prime }{3} \, \Psi  \, G
 \, + \, \dfrac{2 \, f_0}{9} \, \, b_\mu^2 
  \nonumber \\
 & +& \, 8 \, \alpha \, \lambda^2 \, ( \, S \, \Box \, S + G \, \Box \, G \, ) \, - \, 2 \, \lambda \, ( \, S^2 + G^2 \,) \, 
\nonumber \\
& + & 
2 \, ( \beta - f_0^\prime )\, ( \, A \, \Box \, A + B \, \Box \, B \, ) \, - \, 2 \, f_0 \, ( \,A^2 + B^2 \,)
\, - \, 2 \, \lambda \, ( \, \rho  \, \Box \, \rho+ \sigma \, \Box \, \sigma \, ) 
\nonumber \\
& - & 
\,  4  \, \lambda \, f_0^\prime \, ( \,  A \, \Box \, \rho  + \,  B \, \Box \, \sigma \, ) \, + \, 
2 \, \alpha \, ( \, ( \, \Box \, A )^2 + \, ( \, \Box \, B )^2\, ) \, + \, 
8 \, \alpha \, \lambda^2 \, ( \, ( \, \Box \, \rho )^2 + \, ( \, \Box \, \sigma )^2\, )
\nonumber \\
&+& 8 \, \alpha \, \lambda \, ( \, \Box \, A \, \Box \, \rho  + \, \Box \, B \, \Box \, \sigma \, ) .
   \label{quadall}
 \eea  
 The function $\, F(R)$ includes all terms, even nonquadratic,  that depend exclusively on the curvature,   but not on its derivatives. Its specific form   is given by, 
 \bea
 F(R) \, = \,  - \, \dfrac{R}{3} \, f \left( \dfrac{R}{6} \right) \, + \,  \dfrac{\alpha}{108}  \,  {R^3}  
  + \, \dfrac{\beta}{18} \, R^2 \, .
 \label{frrr}
 \eea  
 The constants  $\, f_0, f_0^\prime$ appearing in (\ref{quadall})  stand for $f(0)$ and $f^\prime(0)$ respectively.  
 Since the real function $f(c)$ is arbitrary so is the function  $F(R)$ and hence the constants $\, f_0, f_0^\prime$.  Expanding the function $ f(R/6) $,  the linear in the  curvature term is 
\bea
- \, \dfrac{f_0}{3} \; R . 
\label{rnor}
\eea
This dominates in the weak field  limit but is not a canonically diagonalized Einstein term  $\, - R / 2  $.  However this can be remedied 
by an appropriate  constant  rescaling of the metric, $ g_{\mu \nu} \rightarrow \sigma \, g_{\mu \nu} $, with 
$\sigma = 3 / 2 f_0$, which brings the curvature term in (\ref{rnor})  to its well-known Einstein form $\, - R / 2  $. 

 The fields get mixed in the bilinear terms therefore the mass spectrum is rather difficult to read directly at this stage. Note, especially, the mixing of the curvature with the real part of the  field $F_q$, denoted by $S$.  As has been already discussed,   $F_q$  is dynamical in this formulation since the Lagrangian includes derivatives of it. Isolating the bilinear terms involving the curvature $R$ and the field $S$,  and rescaling the field $S$,  by   $ \, S = \, ( 16 \alpha \lambda^2 )^{ - 1 / 2} \, \hat{S}  $,  so that its kinetic term is canonical, we get  
 \bea
  e^{\, -1} \, {\cal{L}}^{\, (quad)}_{RS}  = &&  F(R) \, +  
  \dfrac{\alpha}{18} \,  R \,  \Box \, R  \,  
  - \, \dfrac{ \, \sqrt{\alpha} }{\, 3} \, \, R \, \Box \, \hat{S}
  \, + \, \dfrac{\, f_0^\prime}{6 \, \sqrt{\alpha}} \,  \, R  \, \hat{S}
  \nonumber \\
   & +& \, 
  \,  \dfrac{1}{2}  \,  \hat{S} \, \Box \, \hat{S} \, - \, \dfrac{1}{8 \alpha \lambda}\,  \hat{S}^2 .
 \label{RS}
 \eea
 The simplest way to derive the tree-level mass spectrum is to find the equations of motion  of all fields involved. For the Lagrangian (\ref{RS}) the equations of motion that follow by varying  the metric  $g_{\mu \nu}$ and  $\, \hat{S}$ are given below. The variations of each term in Eq. (\ref{RS}),  with respect to the metric, are presented in Appendix \ref{app1}. 
 It is essential to note that only the linear terms will be kept in the equations of motion, since we want to find the tree-level mass spectrum, which makes the task much easier. Then for the variations $\delta g_{\mu \nu}$,  given in  Eqs. (\ref{del1}) - (\ref{del4}), we pick only the linear terms. Then  employing the fact that $F(0) = 0$, which follows from Eq. (\ref{frrr}), we arrive at  
 \bea 
 & - & ( \,  R_{\mu \nu} - \dfrac{g_{\mu \nu} }{2} \, \, R \, \; \; ) \, F^\prime(0)
 \nonumber \\
 \, &+& \,  \left(  \, g_{\mu \nu } \, \Box \,  - \nabla_\mu   \,  \nabla_\nu     \,     \right) \, 
 \left( \,  F^{\prime \prime} (0) \; R \, 
 \, + \,   \dfrac{\alpha}{9} \,   \Box \, R \,
  - \, \dfrac{ \, \sqrt{\alpha} }{\, 3} \,   \Box \, \hat{S} \,
 +  \,  \dfrac{\, f_0^\prime}{6 \, \sqrt{\alpha}} \,  \, 
    \, \hat{S} \,  \right)  \; = \; 0 .
 \label{eomgr}
 \eea
 Then we expand $ g_{\mu \nu} $  in the usual manner around the flat metric $ \, n_{\mu \nu} = diag \,  ( \, - 1, + 1, + 1, + 1 \, )  $,  that is $\,  g_{\mu \nu}  \ = \, n_{\mu \nu} + h_{\mu \nu} $. Then,  by defining  the field 
 $\,  \chi_{\mu \nu} $,  see (\ref{chimn}),  and employing   the harmonic gauge, (\ref{har}),  the equation of motion receives the following form
\bea
      -  \dfrac{F^\prime(0)}{2}  \, \,   \Box \, \chi_{\mu \nu}  
 \,  + \,  ( \,  n_{\mu \nu}  \, \Box \, - \,    \partial_\mu   \,  \partial_\nu  \,   ) \,
\left( \, - \, \dfrac{F^{\prime \prime} (0)}{2} \, \Box \, \chi - \, \dfrac{\alpha}{18} \; \Box^{\, 2} \chi  \, \, - \,   \dfrac{ \, \sqrt{\alpha} }{\, 3} \, \Box \, \hat{S} 
\, + \, \dfrac{\, f_0^\prime}{6 \, \sqrt{\alpha}} \, \hat{S} \, 
\right) \, = \, 0 .
\label{eom1}
\eea
The parameters $\,  F^\prime(0) , F^{\prime \prime}(0) $ appearing in this equation depend on $\, f_0 \equiv f(0) , f_0^\prime \equiv f^\prime(0) $ and $\, \beta $ as can be seen from Eq. (\ref{frrr}). The precise relations are given in 
(\ref{zeror}). 
As for the equation of motion that follows by varying  the field $\hat{S}$, this is much easier to be derived leading to
\bea
\Box\, \hat{S} \, - \, \dfrac{1}{4 \alpha \lambda}\,  \hat{S} 
- \, \dfrac{ \, \sqrt{\alpha} }{\, 3} \, \,  \Box \, R
  \, + \, \dfrac{\, f_0^\prime}{6 \, \sqrt{\alpha}} \,  \, R  \, = \, 0 .
\label{eomss}
\eea
Keeping the linear terms in $R$, and in the harmonic gauge, this receives the form
 \bea
\Box\, \hat{S} \, - \, \dfrac{1}{4 \alpha \lambda}\,  \hat{S} 
+ \, \dfrac{ \, \sqrt{\alpha} }{\, 6} \, \,  \Box^{\, 2} \, \chi
  \, - \, \dfrac{\, f_0^\prime}{12 \, \sqrt{\alpha}} \,  \, \Box \, \chi  \, = \, 0 , 
\label{eom2}
\eea
 where $\, \chi =  n_{\mu \nu} \chi^{\mu \nu}$.  Eqs. (\ref{eom1}) and (\ref{eom2}) can be written as 
 \bea
&&   \Box \, \chi_{\mu \nu}  
 \, + \,  ( \,  n_{\mu \nu}  \, \Box \, - \,    \partial_\mu   \,  \partial_\nu  \,   ) \,
 \left( \alpha_1 (  \,  \Box \, \hat{S}    + \Box \, \hat{\Sigma } \, )  + \beta_1 \, \hat{S} + \beta_2 \, \hat{\Sigma} \, \right) \, = \, 0
\label{eom11}
\\  
&& \Box\, \hat{S} \, + \, \Box \, \hat{\Sigma } \, + \, \lambda_1 \, \hat{S} \, + \, \lambda_2 \, \hat{\Sigma} \, = \, 0 ,
\label{eom22}
\eea
where for convenience we have denoted $\,   \hat{\Sigma} \equiv \sqrt{\alpha} \, \Box \, \chi \,  / \, 6  $. 
The constants $\, \alpha_1  , \beta_{1,2}   $,  as well as $\,  \lambda_{1, 2  } $,  can be read from (\ref{eom1}) and (\ref{eom2}).  
Eq. (\ref{eom22}) can be  plugged into (\ref{eom11}) yielding
\bea
\Box \, \chi_{\mu \nu}  
 \, + \,  ( \,  n_{\mu \nu}  \, \Box \, - \,    \partial_\mu   \,  \partial_\nu  \,   ) \,
 \left( \, \gamma_1 \, \hat{S} + \gamma_2 \, \hat{\Sigma} \, \right) \, = \, 0
 \quad , \quad \text{where} \quad \gamma_j = \alpha_1 \, \lambda_j + \beta_j .
\label{eom1111}
\eea
This contracted with the flat metric $\, n^{\mu \nu}$ yields
\bea
\gamma_1 \, \Box \, \hat{S} + \gamma_2 \, \Box \, \hat{\Sigma} \, + \, \dfrac{2 }{ \sqrt{\alpha}  } \, \hat{\Sigma} \, = \, 0 .
\label{eom2222}
\eea
Solving (\ref{eom22}) , (\ref{eom2222}) we get a system of two coupled Klein-Gordon equations, 
\bea
 \Box \, \hat{S}  \, = \, \rho_1 \, \hat{\Sigma}  \, + \, \rho_2 \, \hat{S}
 \quad , \quad 
  \Box \, \hat{\Sigma}  \, = \,  \sigma_2 \, \hat{S} \, + \sigma_1 \, \hat{\Sigma}  \, .
\eea
The constants appearing in this equations can be read from the previous expressions.  Note that 
the ``off-diagonal" coefficients $\rho_1, \sigma_2$ are not equal.  This system can lead to two uncoupled  Klein-Gordon equations by  linearly combining   $\,  \hat{S} , \hat{\Sigma}   $ . In order to implement this we write the above system as 
  \bea
 \Box \,
 \left( \begin{array}{c}
\; \hat{S}    \\
\hat{\Sigma}    \\
 \end{array} \right) 
 \, = \, 
\pmb{M^2} 
 \left( \begin{array}{c}
\; \hat{S}    \\
\hat{\Sigma}    \\
 \end{array} \right) 
 \quad \text{where} \quad 
 \pmb{M^2} \, = \, \left( \begin{array}{cc}
\; \rho_2 &  \rho_1 \,    \\
\, \sigma_2   &  \sigma_1   \\
 \end{array} \right) .
   \label{diagsys}
 \eea
  The system (\ref{diagsys}) can be uncoupled by a real matrix $\, \pmb{A}$ that diagonalizes $\, \pmb{M^2}$, 
  \bea
  \pmb{A} \, \pmb{M^2} \, \pmb{A}^{-1} \, = \, 
  \, = \, \left( \begin{array}{cc}
\; m_1^2 & 0 \,    \\
\, 0   & m_2^2  \\
 \end{array} \right) ,
 \label{diagon}
  \eea
  where $\, m_{1,2}^2$  are the eigenvalues of $\, \pmb{M^2}$. Then the ``rotated" fields $\,  \Phi_1 , \Phi_2$ defined by
   \bea
 \left( \begin{array}{c}
\, \Phi_1    \\
\Phi_2    \\
 \end{array} \right) , 
 \, = \, 
\pmb{A} 
 \left( \begin{array}{c}
\; \hat{S}    \\
\hat{\Sigma}    \\
 \end{array} \right) ,
   \label{diagsysKG}
 \eea
are two  independent Klein-Gordon fields $\, \Phi_1 , \Phi_2$ satisfying
 \bea
 \Box \, \Phi_1 \, - \, m_1^2 \, \Phi_1 \, = \, 0
 \quad , \quad 
  \Box \, \Phi_2 \, - \, m_2^2 \, \Phi_2 \, = \, 0 .
 \label{KG}
 \eea
 The masses squared $\,  m_1^2 \, , \, m_2^2$ are the eigenvalues of the mass matrix defined in  (\ref{diagsys}). They are explicitly given in Appendix \ref{app3}, see (\ref{masses2}),  where we also discuss the  conditions  for them to be  real and nontachyonic.

 The graviton field is given by
 \bea
\xi_{\mu \nu} \, = \, \chi_{\mu \nu}  
 \, + \,  ( \,  n_{\mu \nu}  \, \Box \, - \,    \partial_\mu   \,  \partial_\nu  \,   ) \,
 \left( \, \tau_1 \, \Phi_1 + \tau_2 \, \Phi_2 \, \right) \, .
 \label{grav}
\eea
Since $ \chi^{\mu \nu} $ is transverse so is $ \xi^{\mu \nu} $ that is, $\, \partial_\mu \xi^{\mu \nu} = 0 $, and it 
satisfies the massless Klein-Gordon equation, as can be seen by acting with the  $\, \Box$ operator on $\, \xi_{\mu \nu} $, 
\bea
\Box \, \xi_{\mu \nu} \, = \, 0
\label{grav2}
\eea
if the constants $\,  \tau_{1,2} $ are given by 
\bea
\tau_j \, = \, \dfrac{ ( \gamma_1 \rho_1  -\gamma_2 \,  \rho_2 ) + m_j^2 \,  \gamma_2 }{\rho_1 \, m_j^2  } \quad , \quad
j = 1, 2 .
\label{taus}
\eea
To arrive at (\ref{grav2}),   Eqs. (\ref{eom1111}) and  (\ref{KG}), as well as  (\ref{diagsysKG}) ,  were used. 
Therefore the physical  degrees of freedom of the $\, R, S$  sector are a massless graviton and two massive scalars with masses
given in  (\ref{masses2}).

As for the remaining degrees of freedom, consider the field $\Psi$, which  mixes only with the imaginary part of $F_q$  named $\, G$, in the bilinear terms. The equations of motion are fairly easy to be derived. In particular by varying with respect $\, b_\mu$  one gets
\bea
- \, \dfrac{ 4 \, \alpha}{  9} \,  \, \partial_\mu \, \Box \, \Psi \, - \,  \dfrac{ 4 \, ( \, \beta - f_0^\prime )}{  9} \,  \, \partial_\mu \,  \Psi 
\, + \,   \dfrac{ 8 \, \alpha \, \lambda }{  3} \,  \, \partial_\mu \, \Box \, G \, - \,   
\dfrac{ 4 \,  \lambda \,  f_0^\prime}{  3} \,  \, \partial_\mu \,  G \, + \,  \dfrac{ 4 \,  f_0 }{  9} \,  b_\mu \, = \, 0 ,
\label{varbmu}
\eea 
from which, acting upon it  by  $\,  \partial^\mu $, we get     
\bea
\Box \, \left( - \ \, \dfrac{1}{3} \, \Box \, \Psi \, + \, 2 \, \lambda \, \Box \, G \, \right) \, - \, \dfrac{ \lambda \, f_0^\prime}{   \alpha} \,  \Box \, G  \,
- \,  \dfrac{  ( \, \beta - f_0^\prime )}{  3 \, \alpha } \, \Box \, \Psi \, + \, 
\dfrac{   \,   f_0 }{  3 \, \alpha } \,  \, \Psi \, = \, 0 .
\label{varbmu2}
\eea
On the other hand  the variation with respect $\, G$ yields, 
\bea
- \ \, \dfrac{1}{3} \, \Box \, \Psi \, + \, 2 \, \lambda \, \Box \, G \, 
\, = \, 
 \dfrac{ 1}{  2 \, \alpha} \, G \, - \, \dfrac{ f_0^\prime}{  6 \, \alpha} \, \Psi \, .
\label{varggg}
\eea
Plugging  the left hand side of (\ref{varggg})   into   (\ref{varbmu2})  and by defining 
 $\, \hat{\Psi} \equiv \Psi / 3  $, for convenience,    we get  the system of equations 
 \bea
 \Box \,
 \left( \begin{array}{c}
\; \hat{\Psi}    \\
\hat{G}    \\
 \end{array} \right) 
 \, = \, 
\pmb{m^2} 
 \left( \begin{array}{c}
\; \hat{\Psi}     \\
\hat{G}      \\
 \end{array} \right) 
 \quad \text{where} \quad 
 \pmb{m^2} \, = \, \left( \begin{array}{cc}
\; \delta_{11} & \delta_{12}  \\
\delta_{21}  &  \delta_{22}   \\
 \end{array} \right) .
   \label{diagsys2}
 \eea
The constants appearing in this equations can be read from  previous expressions. 
The ``off-diagonal" coefficients of the mass matrix  $\,  \pmb{m^2} $   are not  equal, in general.   However,  this system  also leads to two independent Klein-Gordon equations, if one diagonalizes the matrix $\,  \pmb{m^2} $  by a real matrix,  
 as we did in the previous case, see  (\ref{diagon}).
 The corresponding masses squared  are the eigenvalues of the mass matrix appearing on the right  of (\ref{diagsys2}). 
 They are analytically given in (\ref{masses33}) where  it is  shown that  they are identical to the masses 
 (\ref{masses}). The reason behind this degeneracy will be discussed later.  
 
 It remains to find the equations of motion for the system of the fields $\, A, \rho$ and $\, B, \sigma$. It is seen from Eq. (\ref{quadall}) that $\, A$ and $\, \rho$ are coupled, but do not mix with $\, B, \sigma$ which are also coupled. Note that  the pertinent Lagrangian terms for the $\, B, \sigma$ system follow exactly  from those of $\, A, \rho$ by replacing $\, A \rightarrow B$ and 
 $\,  \rho \rightarrow \sigma$. Therefore it suffices to study one of these systems. The equations of motion that follow from  (\ref{quadall}), by varying $A$ and $\rho$  respectively, are given below
 \bea
\Box^2 \, ( \, A \, + \, 2 \, \lambda \,  \rho) \,&=& \, \dfrac{( f_0^\prime - \beta )}{\alpha}\, \, \Box \,  A \, + \, 
\dfrac{\lambda \, f_0^\prime  }{\alpha  } \, \Box \, \rho \, + \, \dfrac{ f_0  }{\alpha  } \, A
\label{aro1}
\\
\Box^2 \, ( \, A \, + \, 2 \, \lambda \,  \rho) \, &=& \, \dfrac{ f_0^\prime }{2 \,\alpha}\, \, \Box \,   A \, + \, 
\, \dfrac{1  }{2 \, \alpha  } \, \, \Box \, \rho ,
\label{aro2}
 \eea
 By defining the combination 
 \bea
 Y \, \equiv \, A + 2 \, \lambda \,  \rho ,
 \eea
 and using this to  replace in the equations above the field  $ \, \rho$ in terms of  $\, A \, , Y   $ we get, combining the resulting equations, 
 \bea
 && \Box \, \, Y \, + \, g \, \Box \, \, A \, + \,  \, \, \mu \,  A \, = \, 0
 \nonumber \\
 &&  \Box^2 \, \, Y \, + \, c_1 \, \Box \, \, Y \, + \, c_2 \, \, \Box \,  A \, = \, 0 .
 \label{aro3}
 \eea
 The constants  $\, c_{1,2} , g , \mu $    can be read from Eqs. (\ref{aro1}) and  (\ref{aro2}) and are given in (\ref{stath55}). Acting in the first of (\ref{aro3}) by $\, \Box$ and replacing in the resulting equation  $\, \Box^2 Y$ by the second of  (\ref{aro3}), we get a system 
 which in matrix notation has the following form, 
   \bea
 \Box^{\, 2}  \,
 \left( \begin{array}{c}
\; {A}    \\ 
\, \, Y    \\
 \end{array} \right) 
 \, = \, 
\pmb{ {\cal{M}}^2} 
\, \;  \Box 
 \left( \begin{array}{c}
\; {A}     \\
\, \, Y      \\
 \end{array} \right) 
 \quad \text{where} \quad 
 \pmb{ {\cal{M}}^2}  \, = \, \left( \begin{array}{cc}
\; m_{11} & m_{12}  \\
m_{21}   &  m_{22}   \\
 \end{array} \right)  .
   \label{diagsys4}
 \eea
The mixing  matrix $\,  \pmb{ {\cal{M}}^2} $ is not symmetric, in general.   Its elements are explicitly  given in (\ref{elem}). 
Diagonalizing this, we get two Klein-Gordon equations for some linear combinations of 
$\, \Box \, {A} \, , \, \Box \, Y$. The relevant masses squared are the eigenvalues of the matrix  $\,  \pmb{ {\cal{M}}^2} $ above.  These are found to be identical to (\ref{masses3}), and hence (\ref{masses}).  For a proof see discussion following (\ref{stath55}).
The system of $\, B, \sigma$ has exactly the same mass spectrum as the $ A, \rho  $ system, as we have discussed. 

Before closing this section, we should point out that the masses derived in this section are in a frame in which the  linear in the curvature term  is  $ - f_0 \, R / 3$, see Eq. (\ref{rnor}).  However masses are usually quoted in the Einstein frame in which the curvature term is normalized to $ -  \, R / 2$. As already pointed out, this can be implemented in a trivial manner with a constant rescaling of the metric [ see discussion following Eq.  (\ref{rnor}) ].  The effect of this is  that the masses derived in this section should be multiplied by the factor $\, \sqrt{3 / 2 f_0}$ to derive those in the Einstein frame, which enter  Newton's law. 

To conclude, the mass spectrum consists of  a massless graviton, four scalar degrees of freedom of mass $m_1$ and another four scalars with masses  $m_2$,   given analytically in (\ref{masses2}). The reason for the resulting  mass degeneracy, in this simple model, will be discussed when dealing with the  standard $N = 1 $ supergravity in the Einstein frame, whose spectrum  should coincide with this of the dual theory considered in this section. This task will be undertaken in the following section.

\section{ $N=1\; $ Supergravity} 

The previously defined models are dual descriptions  of standard $N=1\; $ supergravities where  the curvature term has its canonical Einstein form,  $ - \frac{R}{2}$.  In this frame the kinetic and potential terms of the theory  are given by
\bea
e^{-1} \, {\cal{L}}_{kin} \, &=& \,  - {\cal{K}}_{\bar{J} I} \,  \, \partial^\mu \bar{\phi}^{ \bar{J} } \partial_\mu \phi^I\\
e^{-1} \, {\cal{L}}_{pot} \, &=& \, - \, \left( F^I \, F_I - 3 \, e^{\cal{K}}  \, | W |^2  \right) .
\eea
In these $  \phi^I ,  \bar{\phi}^{ \bar{J} }$ denote respectively the scalar fields involved,  and their complex conjugates, and 
$ \, {\cal{K}}_{\bar{J} I}= \frac{\partial^2 {\cal{K}}  }{   \partial \bar{\phi}^{ \bar{J} } \partial \phi^I  }  $.  
The $F$-terms   $ F^I, F_I $ are given by
\bea
F_I \, = \, e^{ {\cal{K}}/2} \, D_I W \, , \,  F^I \, = \, {\cal{K}}^{I \bar{J}}  \, D_{\bar{J}} \, \bar{W}
\quad  \text{where} \quad D_I W  = \partial_I W + K_I W .
\eea
As usual,  the subscripts $ \, I, \bar{I} $ denote differentiation with respect $ \phi^I , \bar{\phi}^{ \bar{I}}$ and $\, K^{I \bar{J}} $ is the inverse of the kinetic matrix $\, K_{\bar{J} I} $.  

For the class of  models studied in the previous section, the superpotential is 
\bea
W = T \Phi + Q C + \Phi \, f(C) , 
\label{super}
\eea
with $f(C)$ an arbitrary  chiral function of $C$,  and  the K\"{a}hler function ${\cal{K}}$  is related to the real function $\, \Omega$ as given in (\ref{KKK}). The latter  is given by 
\bea
\Omega =  \, T + \overline{T} + ( Q \overline{\Phi} \, + \, \Phi \overline{Q}) + 
2\, \alpha \, C \, \overline{C} + 2 \, \lambda \, Q \overline{Q} + 
2 \, \beta \, \Phi \,  \overline{\Phi}  , 
\label{nscale34}
\eea
The complete form of the scalar kinetic part is rather lengthy and will not be presented. However it takes a rather simple form if we keep the bilinear parts in the fields,  and their conjugates,  by expanding ${\cal{K}}_{\bar{J} I}  $ about $ \Phi = C = Q =0 $ preserving  the terms that depend on $T \, , \, \overline{T}$.  The reason behind this expansion relies on the fact that  the point $ \Phi = C = Q =0 $, as we shall see shortly, corresponds to a global minimum of the scalar potential. In this expansion  the kinetic terms are, 
\bea
e^{-1} \, {\cal{L}}_{kin} \, &=& \, 
- \, \dfrac{3}{4} \, \dfrac{ (\partial_\mu ReT )^2  \, + \, (\partial_\mu ImT )^2 }{ ReT^{\, 2} } 
\, + \, \dfrac{3}{ 2\, ReT }\; \left( \, 2 \, \alpha \, \partial_\mu C  \partial^\mu \overline{C}      
\, + \, 2 \, \beta \, \partial_\mu \Phi  \partial^\mu \overline{\Phi}  \right.
\nonumber \\
& &
\left.
\, + \, 2 \, \lambda \, \partial_\mu Q  \partial^\mu \overline{Q} \, + \,   
( \, \partial_\mu Q  \partial^\mu \overline{\Phi} \, + \, \partial_\mu \Phi  \partial^\mu \overline{Q} \, ) \, 
 \right) ,
 \label{quadkin}
\eea
where the first two terms are the ones encountered in the Starobinsky model. Instead of using $Re T$ we can define a real  scalar field $\psi$ by 
\bea
ReT \, = \, - \, f_0 \, e^{ \, \sqrt{ \frac{2}{3} } \, \psi  } .
\label{ret}
\eea 
The choice of the pre-factor of the exponential in (\ref{ret})  is not essential since by  shifting   the field $\psi$ can be changed to anything. However this choice is convenient  since, as we shall discuss,  the minimum of the potential lies at $\, ReT = - f_0$, or same  $\psi=0$.  With this definition the kinetic terms given in (\ref{quadkin}) receive the following form
\bea
e^{-1} \, {\cal{L}}_{kin} \, &=& \, 
- \, \dfrac{1}{2} \, (\partial_\mu \psi )^2 \, - \, \dfrac{3}{4 \, f_0^2} \, e^{ \,- 2 \, \sqrt{ \frac{2}{3} } \, \psi  } \, (\partial_\mu ImT )^2
\, - \, \dfrac{3}{ 2 \, f_0 } \, e^{ \,-  \, \sqrt{ \frac{2}{3} } \, \psi  } \; \left( \, 2 \, \alpha \, \partial_\mu C  \partial^\mu \overline{C}      
\,    \right.
\nonumber \\
& &
\left.
+ \, 2 \, \beta \, \partial_\mu \Phi  \partial^\mu \overline{\Phi}
\, + \, 2 \, \lambda \, \partial_\mu Q  \partial^\mu \overline{Q} \, + \,   
( \, \partial_\mu Q  \partial^\mu \overline{\Phi} \, + \, \partial_\mu \Phi  \partial^\mu \overline{Q} \, ) \, 
 \right) .
 \label{quadkin2}
\eea
Expanding the exponentials about the point $\psi = 0$, anticipating the fact that at the minimum $\psi = 0$ , we get
\bea
e^{-1} \, {\cal{L}}_{kin} \, &=& \, 
- \, \dfrac{1}{2} \, (\partial_\mu \psi )^2 \, - \, \dfrac{3}{4 \, f_0^2} \, (\partial_\mu ImT )^2
\, - \, \dfrac{3}{ 2 \, f_0 } \,  \left( \, 2 \, \alpha \, \partial_\mu C  \partial^\mu \overline{C}      
\,    \right.
\nonumber \\
& &
\left.
+ \, 2 \, \beta \, \partial_\mu \Phi  \partial^\mu \overline{\Phi}
\, + \, 2 \, \lambda \, \partial_\mu Q  \partial^\mu \overline{Q} \, + \,   
( \, \partial_\mu Q  \partial^\mu \overline{\Phi} \, + \, \partial_\mu \Phi  \partial^\mu \overline{Q} \, ) \, 
 \right) .
 \label{quadkin3}
\eea
These kinetic terms are not canonically normalized. Moreover they mix in the $Q, \Phi$ sector. 
The kinetic mixing matrix associated with  the $Q, \Phi$ sector has determinant proportional to $\, 4 \beta \lambda -1 $.  Therefore when either $\beta$ or $\lambda $ vanish, that is when  there are no  diagonal quadratic kinetic terms for either $Q$ or $\, \Phi$ fields,  one of its eigenvalues is negative. This signals the appearance of  ghost states !  Taking $ 4 \beta \lambda > 1  $ all eigenvalues of the kinetic mass matrix in (\ref{quadkin3}) are positive definite and  this is a necessary condition in order to avoid ghosts. 
Note that this condition lies in the range where the mass squared of the dual theory are positive definite, see Eq. (\ref{cond3}).

Having discussed the kinetic part we now move on to study the scalar potential. 
The complete potential has the following form, 
\bea
V \, &=& \, \dfrac{ 9 }{ 2 \lambda \, ( 4 \beta \lambda - 1 \,) \, \Omega^2  } \, 
 \cdot \,\Bigg( \, ( | \, C - 2 \lambda \, ( T +f(C) ) \, |^{\,2} +  ( \,4 \beta \lambda - 1 \,) \, \Big( \, | C |^{\,2} + \dfrac{\lambda}{\alpha} \,   | \, \Phi f^\prime(C) + Q  \, | ^2 
\nonumber \\ 
& &
 + \,   2 \lambda \, ( \, C Q \overline{\Phi} + H.c. \, )  \,   +  2 \lambda  \, ( \, T  + 2  f(C)  -  C f^\prime(C) + H.c. \, ) \, \, | \Phi |^2 \, \Big)
 \Bigg) .
  \label{potit} 
\eea 
The first three terms are manifestly positive definite. The last two are not and the potential in not bounded from below.  Additional terms need be introduced to stabilize the scalar potential, as the ones employed in \cite{EKN,KALLOSHp1}. Adding a single stabilizing term \cite{KALLOSHp1}
\bea
-  \, \zeta \, | \Phi |^4
\label{kallosh}
\eea
to the function $\Omega$ is adequate to stabilize the scalar potential as we will discuss. That done, the scalar potential receives a rather complicated form, which, however, we can handle analytically. Its complete expression is given by  (\ref{potall}) and (\ref{pfun}). In Appendix \ref{appot} we discuss in detail  its minima and its stability.  In fact we find that the potential has a minimum at the point
\bea
ReT = - f_0, \; ImT = 0, \, C = \Phi = Q = 0 .
\label{extrem}
\eea
As we show in Appendix (\ref{appot}) there are values of $\zeta$ for which the potential is positive definite for any value of the fields involved. Therefore this  minimum   is actually the absolute minimum of the potential. 
At the minimization point (\ref{extrem}) the scalar potential vanishes, see (\ref{pderp}), that is  we have a Minkowski vacuum. Moreover at this vacuum  supersymmetry remains unbroken, since $ \vev{F_I}  = 0$ for any value of $I$. 

Being the absolute minimum, the matrix of the second derivatives, at this point, should be positive, i.e. all of its eigenvalues should be positive definite. This statement is equivalent to saying that there are no tachyonic masses in the spectrum of scalars, which we shall prove in the following. We point out that the mass spectrum is independent of the stabilizer.  Actually, expanding  the potential about the point  (\ref{extrem}) its  quadratic terms do not depend on $\zeta$ as can be seen from the form of the potential, given in (\ref{potall}),  using the fact that at the minimum (\ref{pderp})  holds. 
This differs from other $R^2$ supergravity models,  in which the scalaron has a $\zeta$-dependent mass due to the fact that the lowest minimum of the potential  is  $\zeta$-dependent, as well \cite{addazi}.  

In fact by expressing  the fields in terms of their real and imaginary parts, and trading   $ReT$ for  the field $\psi$, defined in Eq. (\ref{ret}), we find that the quadratic terms arising from the potential are given by 
\bea
e^{-1} \, {\cal{L}}_{pot}^{(quad)} \, &\equiv & \, - \, V_{quad} \,    
\nonumber \\
&=& - \, k_1 \, \Big( 
3 \, \lambda \, ( Im T )^2 \, + \, 3 \, ( \, \beta + ( f_0^\prime )^2 \lambda -f_0^\prime \,  ) \, ( Im C )^2 \,  
+ \, + 3 \, ( 2 \lambda f_0^\prime - 1) \, Im T  \, Im C
\Big.
\nonumber \\ 
& &
\, \Big. 
\quad \quad \; \;  + \, ( \, 2 \lambda f_0^2 \, ) \, \psi^2 + 3 \, ( \, \beta + ( \, f_0^\prime )^2 \lambda -f_0^\prime \, ) \, ( Re C )^2 \,  
+ \, \sqrt{6} \,f_0 \,   ( \, 1 - 2 \lambda f_0^\prime \, ) \, (ReC) \, \psi \Big)
\nonumber \\
&  & 
- \, k_2 \, \Big( \, ( Re Q )^2 \, + \, ( \, ( f_0^\prime )^2 + 4 \alpha \, f_0 \, ) \, ( Re \Phi )^2 \, + \, 2 \, f_0^\prime \, ( Re Q ) ( Re \Phi ) \Big.
\nonumber \\
& & 
\Big.
\quad \quad \quad + ( ReQ \rightarrow Im Q , Re\Phi \rightarrow Im \Phi ) 
\Big) .
\label{quadpot}
\eea
In it the constants $\, k_{1,2}$ are given by 
\bea
 k_1 = \dfrac{ 3 }{ 2 \, f_0^2 \, ( 4 \beta \lambda -1 ) } \, \quad k_2 = \dfrac{9}{8 \alpha f_0^2} \,  .
 \label{kkk1} 
\eea
In (\ref{quadpot}) the fields are mixed pairwise. In fact $ \psi  $ mixes with $\, Re C$,  $ Im T  $ mixes with $\, Im C$, 
$ ReQ   $ mixes with $\, Re \Phi$ and $ ImQ   $ mixes with $\, Im \Phi $.  Having the bilinear kinetic and potential terms it is fairly easy to find the mass spectrum. This task is facilitated a great deal by the fact that the mixings among the fields are done in a pairwise manner and we only have to generalize two by two matrices. Due care should be taken by the fact that  in the kinetic part the fields  are not canonically normalized and,  besides,  mixings occur in the $Q, \Phi$ sector, as is evident from   (\ref{quadkin3}).  
That done we find that the masses are exactly the same with the ones derived in the dual theory if the latter are multiplied by a factor $\, \sqrt{3 / 2 f_0}$.  The origin of this difference was adequately 
explained in the concluding remarks of the previous section, and is due to the fact that  masses read in the Einstein frame differ by a constant from those in other frames in which the curvature term appears with  a different normalization.

An alternative, and perhaps more elegant way, to deal with the mass spectrum, and also shed light to the issue of mass degeneracy, is to change the superfield basis. 
Concerning the mass degeneracy, a double mass  degeneracy is expected among the scalars due to supersymmetry that is not broken at the minimum of the potential. Scalar degrees of freedom have same masses with their fermionic counterparts,  the latter occurring in two helicity states. Therefore, to each Weyl fermion, there corresponds  two real scalar fields having the same mass.  However a larger mass degeneracy is observed, actually twice the one expected.   In order to treat the system in a more symmetric manner  and  find the source of the degeneracy,  we had better change the superfield basis, working instead  with  shifted fields, defined by
\bea
T^\prime = T + f_0 \quad , \quad Q^\prime = Q + f_0^\prime \Phi .
\eea
These shifts are dictated by the form of the superpotential (\ref{super}),  when its last term $\, \Phi f(C)$  is expanded in powers of 
$C$, which in this way receives the following form, 
\bea
W = ( T + f_0 ) \, \Phi + (  Q + f_0^\prime \Phi ) \, C + {\cal{ O}} ( \Phi C^2 ) .
\label{super2}
\eea
The last term in the expression above is at least  cubic in the superfields involved and will not actually concern us. The kinetic function $\, \Omega$ given in Eq. (\ref{nscale34}) can then be expressed in terms of the new multiplets. To that purpose, it  proves easier to use a rescaled field,  $\, \Phi^\prime = 2 \sqrt{ \alpha \, f_0 } \, \, \Phi $. Within $\, \Omega$  mixings of  $\Phi^\prime, Q^\prime $ occur,   and by a suitable orthogonal rotation $ \pmb{R}$ of the  $\Phi^\prime, Q^\prime $ superfields these can be uncoupled 
 \bea
 \left( \begin{array}{c}
\; {\Phi^\prime}    \\
Q^\prime    \\
 \end{array} \right) 
 \, = \, 
\pmb{R} \; 
 \left( \begin{array}{c}
\; {\Xi_1}     \\
\Xi_2     \\
 \end{array} \right) 
 \quad \text{where} \quad 
 \pmb{R} \, = \, \left( \begin{array}{cc}
\; cos \theta & sin \theta  \\
 - sin \theta   & cos \theta   \\
 \end{array} \right) ,
   \label{rotbas}
 \eea
leading to the following  $ \Omega, W  $  functions,
\bea
\Omega \; &=& \, -2 f_0 + \sqrt{2 f_0   } \,  ( t + \overline{t}) + \Sigma_1 \overline{\Sigma_1} + \Sigma_2 \overline{\Sigma_2}  + \Sigma_3 \overline{\Sigma_3}  
\nonumber \\
W \, & =&
 \, (  t \, , \, \Sigma_3  ) \, 
 \pmb{R} \, \left( \begin{array}{c}
{m_2 \, \Sigma_2}    \\
m_1 \, \Sigma_1    \\
 \end{array} \right) + E .
\label{wwww}
\eea
To cast these functions as above, we have also implemented the following trivial  rescalings
\bea
T^\prime = \sqrt{2 f_0   } \, \, t \; , \; \Xi_1 = ( \sqrt{2 \alpha} \,  m_2 ) \, \Sigma_2 \; , \;  \Xi_2 = ( \sqrt{2 \alpha} \,  m_1 ) \, \Sigma_1
\; , \; C = \dfrac{1}{\sqrt{ 2 \alpha } } \,\Sigma_3 ,
\eea
where $m_{1,2}$ are exactly the masses given in Eq. (\ref{masses2}).  The last term $E$ in the superpotential $W$ is a function of 
$   \Sigma_1, \Sigma_2, \Sigma_3$ which is at least  cubic in the superfields.  This will not be  explicitly shown, since will not concern us for the discussion that follows.  It suffices to say that  it has the form    $ ( a \, \Sigma_1 + b \, \Sigma_2 ) \, R(\Sigma_3)$ with   $R(\Sigma_3)  $  a function  at least quadratic in the superfield $\, \Sigma_3$. 
The advantage of working with the basis of superfields $\, t , \Sigma_i$  is twofold. The first is that the function   $\Omega$ is brought to a form corresponding to a K\"ahler potential whose scalar fields parametrize the coset space 
$ SU(4,1) / SU(4) \times U(1)$. Such a parametrization is a general feature of the no-scale models. The scalar kinetic terms are those of  a nonlinear sigma model having as isometry group  the noncompact $  SU(4,1)$ symmetry. The second reason is that,  the scalar fields corresponding to the $ t \, , \, \Sigma_i $ multiplets have vanishing values at the absolute minimum of the potential, which, as we have already said, is a Minkowski vacuum with unbroken supersymmetry.  
This, in conjunction with the fact that the superpotential is at least quadratic in the fields,  has the effect that  the only quadratic terms of the potential, when it is expanded about its Minkowski vacuum, are those stemming from the  $F$-terms. In  particular, one needs only to calculate  the derivatives of the first terms in the superpotential given in Eq. (\ref{wwww}),  and the last term  $E$ plays no role in  the mass spectrum.  This facilitates the calculation a great deal,  in both  identifying the scalar mass eigenstates, and find the  mass spectrum, and also tracing the source of the mass degeneracy.  In particular, in this basis  the K\"ahler metric, and its inverse,  receive a  simple diagonal form, at the minimum,  
\bea
{\cal{K}}_{ \overline{J} I} \, = \, \dfrac{2 f_0}{3} \, \ \delta_{ \overline{J} I}
\quad , \quad  {\cal{K}}^{I \overline{J} } \, = \, \dfrac{3}{2 f_0} \, \ \delta^{I  \overline{J} } ,
\label{kkkaaa}
\eea
and thus the quadratic terms of  the potential are given by 
\bea
V_{quad} \, &=& \, e^{ \cal{K} } \, \left(  \dfrac{3}{2 f_0} \right)  \, \sum_I \, | \partial_I W |^2 
=
 \left( \dfrac{3}{2 f_0} \right)^2 \, \left( \,  m_1^2 \, ( | x_1 |^2 + | \sigma_1 |^2  ) 
+ m_2^2 \, ( | x_2 |^2 + | \sigma_2 |^2  ) \, \right) .
\label{pot22}
\eea
In this $\, \sigma_{1}$ and $\, \sigma_{2}$ are the scalars of the multiplets $\Sigma_{1,2}$ respectively,  while $\, x_{1,2}$ are those of the rotated  multiplets defined by $\,  X_1 = c \, \Sigma_3 + s \, t  \, $ and  $\, X_2 = -s \, \Sigma_3 + c  \, t  $. 
Note that $X_{1,2}$ are exactly the combinations of  $t, \Sigma_3$ multiplets appearing in the first part of  the superpotential $W$ given in  Eq. (\ref{wwww}). As for the kinetic terms, collecting the quadratic terms, using the fact that the K\"ahler metric in the $t, \Sigma_i$ basis is diagonal having the simple form (\ref{kkkaaa}), we get, after replacing the scalars  $t, \sigma_3$  by $x_{1,2}$, 
\bea
e^{-1} \, {\cal{L}}_{kin} \, = \, 
- \left(  \dfrac{3}{2 f_0} \right)  \, \left( \, | \partial_ \mu x_1 |^2 + | \partial_ \mu x_2 |^2 + | \partial_ \mu \sigma_1 |^2 + | \partial_ \mu \sigma_2 |^2 \right) .
\label{kin2}
\eea
From Eqs. (\ref{pot22}) and (\ref{kin2})  we see that  $\, x_1, \sigma_1$  have common  masses squared $\,  (  {3}/ {2 f_0} )  \, m_1^2  $ and  $\, x_2, \sigma_2$  have $\,  (  {3}/ {2 f_0} )  \, m_2^2  $, with $m_{1,2}^2$ given in (\ref{masses2}). This 
we have already found previously in an alternative manner.  
 
 Note  that by working in the new basis not only  the scalar sector is treated more symmetrically, but also   the 
  flat limit, $ M_{Planck} \rightarrow \infty$,  is more easily obtained. In fact, 
modulo normalization,  the scalar  kinetic and the mass terms given before  are those encountered in a globally supersymmetric model 
involving four chiral multiplets.    
  In particular, in the flat limit  only renormalizable couplings survive, and the model smoothly goes to  that of a a rigid supersymmetry with four (normalized) chiral fields, given by
 \bea
  \hat{ X }_i =    \sqrt{ \dfrac{3}{2 f_0}  } \, X_i \quad , \quad 
    \hat{ \Sigma }_i =    \sqrt{ \dfrac{3}{2 f_0}  } \, \Sigma_i 
    \quad , \; i = 1,2 ,
 \eea
 and a superpotential 
 \bea
 \hat{W} = \hat{m}_1 \, \hat{ X }_1 \, \hat{ \Sigma }_1 + \hat{m}_2 \, \hat{ X }_2 \, \hat{ \Sigma }_2 + \text{  ``cubic terms." }
 \label{hatw}
 \eea
In this,  $ \hat{m}_i  $   are the rescaled masses $ \hat{m}_i =  \sqrt{{3}/{2 f_0}  } \, m_i$,  and in this limit  only  
the renormalizable cubic terms have been kept in the $E$-term appearing in Eq. (\ref{wwww}),  designated as ``cubic terms" in Eq (\ref{hatw}). The scalar potential of the theory is that of a global supersymmetry,  having the well-known  form
\bea
V \, = \, \sum_{i=1}^2 \, \left( \, \left|  {\partial \, \hat{W}}/{\partial \hat{ X }_i} \right|^2 + \left|  {\partial \, \hat{W}}/{\partial \hat{ \Sigma }_i} \right|^2 \, \right) ,
\eea
whose vacuum does not break supersymmetry. Then, from this form it is evident that to each multiplet  pair 
$  \hat{ X }_i \, , \, \hat{ \Sigma }_i$ their corresponding  complex scalars, i.e., four degrees of freedom,  have common masses 
$\, \hat{m}_i  $ appearing in the superpotential $\hat{W}$. This degeneracy is due to the specific choice for the superpotential, given in  (\ref{hatw}), whose  only mass terms  are those mixing  $  \hat{ X }_i \, , \, \hat{ \Sigma }_i$, with a  mass parameter $\, \hat{m}_i  $, in the way shown above. 
 Had we included the fermionic components we would have  found that the  fermionic Weyl components of 
 $  \hat{ X }_i \, , \, \hat{ \Sigma }_i$ multiplets also mix in their mass terms, exactly in the same manner, with a  mass parameter         $\, \hat{m}_i  $. Therefore they  compose the  left- and right-handed components of a Dirac fermion having mass $\, \hat{m}_i  $. Thus, we have  four fermionic degrees of freedom, of mass   $\, \hat{m}_i  $,  which exactly match the bosonic degrees of freedom,  as expected in unbroken supersymmetry.  

\begin{figure}[h]
\centering
  \centering
  \includegraphics[width=.6\linewidth]{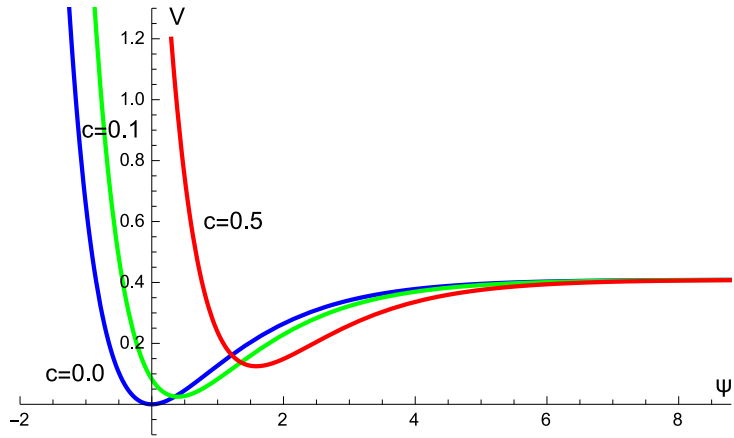}  
\caption{The potential (\ref{potdir}) as function of $\psi$ for representative  real values of the field $ C $ denoted by $c$ in the Figure.  The  curves displayed correspond to  $c =0.0 , \, 0.1 \,$ and 
$c = \, 0.5$ as labeled. 
The couplings have been taken  $\alpha=0.1, \, \beta=3.0 ,$ and $ \, \lambda = 1.0  $,  and the 
function $f(C)$ is $\,  f_0 + f_1 C + f_2 C^2$ with  $f_0 = 0.5, \, f_1= 2$  and $ f_2 = 0.1$.  
The remaining fields have been fixed to their minimum values ( see Eq. \ref{extrem}).
}
\label{fig1}
\end{figure}

\begin{figure}
 \vspace*{5mm}
 \centering
   \includegraphics[scale=.6]{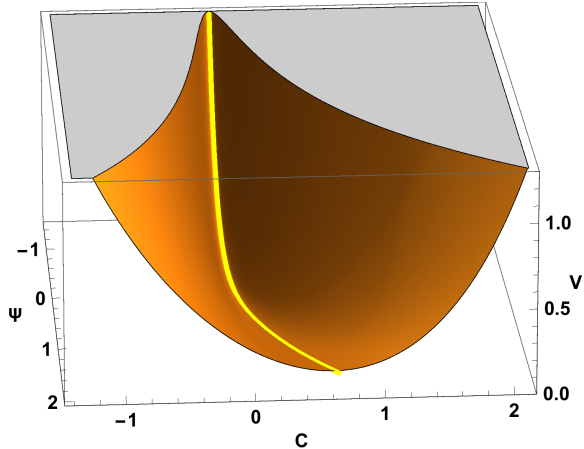} 
   \caption{The  potential (\ref{potdir}) as function of $\psi$ and $\, c = Re C$.   The couplings and the function $f(C)$ are as in 
   Fig.~\ref{fig1}.  The curve drawn on the surface of the potential  is the  trajectory defined by $\, \frac{ \partial V}{ \partial Re C } \, = \, 0 $ .  As in Fig. \ref{fig1} the remaining fields have been fixed to their minimum values.
   }
   \label{fig3}
  \end{figure}

Having found the complete form of the potential, and in order to have a better understanding of its behavior, we start from its absolute minimum and move in the $T, C$ direction, keeping the remaining fields to their vacuum values, $\, \Phi = Q =0$, see (\ref{extrem}) . 
We then see, from (\ref{potall})  and (\ref{pfun}), that  it has the simple  form,   
\bea
V( T, C ) \, &=& \, \dfrac{ 9 }{ 2 \lambda \, ( 4 \beta \lambda - 1 \,) \, (  T + \overline{T} + 2 \alpha | C |^2   )^2  } \, 
\nonumber \\ 
& &
  \cdot \,\Big( \, ( 4 \beta \lambda - 1 \,) \, | C |^{\,2} + | C - 2 \lambda \, ( T +f(C) ) |^{\,2} 
  \Big) .
\label{potdir}    
\eea  
Various profiles of the potential are presented in Fig. \ref{fig1}, for some  representative choices of the function $f(C)$.   In this figure we display the potential as function of $\psi$, defined in (\ref{ret}) by   
$ ReT \, = \, - \, f_0 \, e^{ \, \sqrt{ \frac{2}{3} } \, \psi  } $, 
 for various values of $c \equiv Re C$ and $Im C =0$. In each case shown, the potential has the form of a Starobinsky-like potential  whose minima  lie higher, for higher values of $c$,  the lowest minimum being attained  for $c = 0$. 
 The potential as function of $\, \psi$ and $Re C$,  with the remaining fields set to zero, has a funnel-like shape, as  shown in Fig.  \ref{fig3}, which narrows as $\psi$ gets smaller. On this curve we have also drawn the trajectory of the minima in the $Re C$ direction for fixed $\psi$ values, i.e. the points corresponding to $\frac{\partial V}{\partial Re C} = 0$.  The role of  this trajectory will be discussed in later. 
 
For vanishing values  of $C$  the potential (\ref{potdir})  takes the simple form, 
\bea
V( T, C = 0 ) \, = \, \dfrac{ 18 \, \lambda  }{ (\, 4 \beta \lambda - 1\, )  } \, 
\dfrac{  | \,T + f(0) \, |^2}{   (  \, T + \overline{T} \,  )^2 } \,= \,
 \dfrac{ 9 \, \lambda  }{ 2 \,(\, 4 \beta \lambda - 1\, )  } \, 
 \dfrac{  ( \,Re T + f(0) \, )^2 \, + \, ( Im T )^2}{   (  \, Re T \,  )^2 }     ,
\eea
which by replacing  $Re T$ by the field $\, \psi$, using (\ref{ret}), and in the particular  direction $Im T = 0$,
 it receives the well-known form of  the one field Starobinsky potential, 
\bea
V(\psi) \, = \,  \dfrac{ 9 \, \lambda  }{ 2 \,(\, 4 \beta \lambda - 1\, )  } \, 
( \, 1 - e^{ - \,\sqrt{ \frac{2}{3} } \, \psi}    \,)^{\, 2}
\, \equiv \,  \dfrac{3 \, \mu_s^2 }{  4  } \, ( \, 1 - e^{ - \,\sqrt{ \frac{2}{3} } \, \psi}    \,)^{\, 2} .
\label{staro22}
\eea
Its scale $\, \mu_s$  is set by a combination of  the parameters $\lambda, \beta$,  as is evident from  (\ref{staro22}). 

Therefore,   we have constructed  a higher derivative  supergravity of the form  $F(R, R \Box R)$, whose dual  Einstein-Hilbert description  has no ghosts  and its potential  is described by four complex scalar degrees of freedom.  This is positive definite,  having one stable Minkowski  vacuum, with unbroken supersymmetry. Starting from this minimum, and moving in  particular directions, the scalar potential has the shape of  the  well-known Starobinsky model. It is mainly for this reason that we dubbed the class of models described in this work as deformed Starobinsky models. We are aware that the virtues of the single-field  Starobinsky model, which  successfully describes  cosmological inflation in a simple manner, may not be shared by the class of models considered here. In fact the models considered here are unavoidably multiscalar and only  in particular directions have  profiles reminiscent of the Starobinsky potential. However this by itself is not adequate to reach the conclusion that they are successful in describing cosmological inflation. 

A thorough study of the cosmological consequences of these models  are beyond the scope of the present paper. However  we may present reasons why this class of models  may be  of relevance for  cosmological inflation. 
Although it is premature to reach  definite conclusions,   the qualitative features of the potential are such that  a two-field inflation may be sustained, in principle.  In particular, starting from some initial values of the fields 
$\psi, Re C$, freezing the remaining degrees of freedom to their vacuum values,   the field $Re C$ rolls down to reach the  minimum in the $Re C$-direction, that is it starts tracking the   trajectory  drawn in Fig. \ref{fig3}. Then the  fields   $Re C$ ,$\psi$ are decreased continuously, in their journey to the  absolute  minimum, which   corresponds to  $\psi = 0 , Re C = 0$.   Although we have verified numerically this behavior,  evidently this scenario has to be taken with a grain of salt, as far as its cosmological predictions are concerned,  and a thorough investigation  will appear in a forthcoming publication.  In particular the role of the other fields, that we have frozen at their minimum values, may upset the whole picture since during the evolution the frozen scalar degrees of freedom  may depart from their minimum values destabilizing in this way  the inflationary trajectory.  

Concluding this section, we have seen that the departure from the linearity of the $Q$-field, that plays the role of  a Lagrange multiplier when it linearly appears in the theory, besides leading to a standard $N=1$ old-minimal supergravity  theory free of ghosts, it also  yields supergravity models having a no-scale structure,  with K\"ahler potentials describing  the  coset  $SU(4,1)/SU(4) \times U(1)$.   These are characterized by a stable scalar potential, described by four complex fields, with a single Minkowski vacuum and unbroken supersymmetry. In particular directions the scalar potential has a Starobinsky-like form.  A detailed study of its
cosmological consequences  will be presented in a forthcoming publication.

  \section{Discussion - Conclusions}

  In this work we have addressed the question whether higher derivative supergravity models can be dual to  ordinary ghost-free $N=1$ supergravities. It has been long known that such a duality, for $F(R)$-supergravities, can be established  with the aid of two pairs of chiral multiplets, one pair serving  as Lagrange multiplier and the other  solving in terms of  $\mathcal{R}$ and $T(\mathcal{R})$ respectively.  The higher derivative supergravities constructed in this manner do not include derivatives of the scalar curvature, while introducing  additional multipliers leads  to more general theories  involving, also, derivative  terms  
 $\Box R, \Box^2 R  $ etc. \cite{CECOTTI}.  It is also known, that this description suffers from the appearance of  ghost states (poltergeists) in the Einstein frame $N=1$ supergravity, which are in general as many as the Lagrange multipliers, which should decouple from the spectrum. 
 
 The aforementioned construction can be generalized by modifying the kinetic function $\Omega$ so as to include genuine kinetic terms, for the chiral multiplets involved to implement the duality.  In this approach we managed to get ghost-free supergravity models, in the Einstein frame, which have dual formulation as higher derivative supergravities. Using the minimal construction,  with the least number of chiral fields present, the characteristics of these theories is that in addition to curvature dependent terms,  $\Box R$-terms  are also present. Besides, since in this formalism one of the  Lagrange multiliers is promoted to a dynamical field,  the  higher derivative supergravity remains coupled to this chiral field which, unlike previous constructions, is  not eliminated from the action. In this way the dual theories have the same number of dynamical degrees of freedom.  
 Within  this framework we worked out specific examples showing analytically  the coincidence of the mass spectrum between the two descriptions and the absence of ghost states.   
 
 The construction presented in this work, although it shares features of  existing descriptions found in the literature,  is different from them in many respects. For instance,  auxiliary fields  are also present in the case of the  constrained superfield formalism \cite{ROCEK, KOMAR, ANTODUDAS, KUZENKO, farakos1, farakos2, mosk} but obviously in our work the fact that they  survive in the on-shell Lagrangian is not due to any constraint.  
It is also known that other higher derivative models,  including operators of the form  $\sim \mathcal{D}\Phi\mathcal{D}\Phi\overline{\mathcal{D}}\Phi^{\dagger}\overline{D}\Phi^{\dagger}$, can lead to a ghost-free theory in four dimensions \cite{DudasG,ovrut1, ovrut2, ovrut3}.  
The inclusion of these terms leads to cubic equations for the auxiliary fields of the chiral multiplets,  without inducing kinetic terms for them. The solution of these equations, for the elimination of the auxiliary fields,  gives three inequivalent on-shell theories.  Restrictions from the effective field theory rules out the two of them, leaving only one consistent solution \cite{louis}.  In our approach the couplings of all chiral  fields involved, in the higher-$R$ description,  arise  naturally,  while in the Einstein frame the model is
quite conventional.

The simple higher derivative theories considered in this work lead unforcefully to generalizations of the supersymmetric Starobinsky model, in the $N=1$ supergravity formulation, in the Einstein frame. These models are of the  no-scale type,  with the associated 
K\"ahler function  having the structure of the  $SU(4,1)/SU(4) \times U(1)$ coset manifold. 
We have presented a preliminary discussion concerning  the  possibility that  the resulting scalar potential of these models can drive cosmological inflation.  Recall that  in the description of $F(R)$-supergravity in the Einstein frame, although  ghosts may  decouple from the theory \cite{CECOTTI} the potentials arising are in general unstable or, in special cases,  not obviously leading to inflationary behavior \cite{DLT}.  In the model worked out in this work,  the arising  potential  features  interesting properties. It exhibits Starobinsky directions,  as in the supersymmetric completion of 
$R+R^2$ theory, and besides, no extra instabilities are developed.  In particular, although in our case the scalar potential is described by  four complex scalar fields, nevertheless introducing a normalizing term  \cite{KALLOSH} is adequate to render the potential stable, exhibiting an absolute Minkowski vacuum that does not break supersymmetry. 
 The cosmological evolution may, in principle,  lead  to a multifield inflation.   
We find this case interesting enough to be considered, although additional stabilizing terms, if introduced, may lead to the conventional single-field inflation.  
 A systematic study of the cosmological aspects of this kind of  models is under study and the results will appear in a forthcoming publication.

 \vspace*{4mm}  
{\textbf{Acknowledgements}}   

This research has been financed by NKUA ( National Kapodistrian University of Athens ).  A.B.L.  wishes to thank K. Tamvakis and A. Kehagias  for illuminating discussions. 
  
 \appendix
 \section{Variation with respect to the metric} \label{app1}
 
 The following variation formulae are useful towards deriving results given in the main text. 
 One finds that 
 \bea
 \delta \int \, \sqrt{g} \, F(R) \, = \,  \int  &  \sqrt{g} \,  & 
\Bigl\{ \,
 \dfrac{{g^{\mu \nu} }}{2} \, F(R)  - R^{\mu \nu} \, F^\prime(R) 
 \, + \, \left(  \, g^{\mu \nu } \, \Box \, R - \nabla^\mu   \,  \nabla^\nu   R  \,     \right) \,  F^{\prime \prime} (R) 
 \Bigr.
 \nonumber \\  
 &&
 \, + \, 
 \Bigl.
\left(  \, g^{\mu \nu}  ( \, \nabla \, R  )^2 - \nabla^\mu  R  \,  \nabla^\nu   R  \,     \right) \,  F^{\prime \prime \prime} (R) \, 
\Bigr\} \, \delta g_{\mu \nu} \,
\label{del1}
 \eea
 and also 
 \bea
 \delta \int \, \sqrt{g} \, R \, \Box \, R \, = \,  
  \int &  \sqrt{g} \,  & 
\Bigl\{ \,
 - \, \dfrac{{g^{\mu \nu} }}{2} \, ( \, \nabla \, R  )^2 + \nabla^\mu  R  \,  \nabla^\nu   R
 \, - \, 2 \, R^{\mu \nu} \, \Box \, R
 \Bigr.
 \nonumber \\
 &&
 \, + \, 2 \, ( \,  g^{\mu \nu}  \, \Box \, - \,    \nabla^\mu   \,  \nabla^\nu  \,   ) \, \Box \, R \,
 \Bigl.
\Bigr\} \, \delta g_{\mu \nu} \, .
\label{del2} 
 \eea
 Moreover
 \bea
 \delta \int \, \sqrt{g} \, R \, \Box \, \hat{S} \, = \,  
  \int &  \sqrt{g} \,  & 
\Bigl\{ \,
 - \, \dfrac{{g^{\mu \nu} }}{2} \,  \nabla_\lambda \, R \, \nabla^\lambda \hat{S}  + \nabla^\mu  R  \,  \nabla^\nu   \hat{S}
 \, -  \, R^{\mu \nu} \, \Box \, \hat{S}
 \Bigr.
 \nonumber \\
 &&
 \, + \,  ( \,  g^{\mu \nu}  \, \Box \, - \,    \nabla^\mu   \,  \nabla^\nu  \,   ) \, \Box \, \hat{S} \,
 \Bigl.
\Bigr\} \, \delta g_{\mu \nu} \,
 \label{del3}
 \eea
 and 
  \bea
 \delta \int \, \sqrt{g} \, R \, \hat{S} \, = \,  
  \int &  \sqrt{g} \,  & 
\Bigl\{ \,
( \,  R^{\mu \nu} - \dfrac{g^{\mu \nu} }{2} \, R \, ) \, \hat{S} 
 \, + \,  ( \,  g^{\mu \nu}  \, \Box \, - \,    \nabla^\mu   \,  \nabla^\nu  \,   ) \, \hat{S} \,
\Bigr\} \, \delta g_{\mu \nu} \, .
 \label{del4}
 \eea

 \section{Expansion about the flat metric} \label{app2}
 
 For the expansion about the flat space-time  metric, the pertinent formulae are given below. 
 \bea
   g_{\mu \nu}  \ = \, n_{\mu \nu} + h_{\mu \nu}  \quad \text{where} \quad   n_{\mu \nu} = diag \,  ( \, - 1, + 1, + 1, + 1 \, )
 \eea
 and one defines, in the usual manner,  
 \bea
 \chi_{\mu \nu} \equiv h_{\mu \nu} - \dfrac{n_{\mu \nu} }{ 2 } \,  h  \quad \quad {\text{where}} \quad h \equiv  n_{\mu \nu} h^{\mu \nu } .
 \label{chimn}
 \eea
In the harmonic ( de Donder ) gauge we have 
\bea
\chi_\nu  \equiv \partial_\mu  \, \chi^{\mu}_{ \nu} =0 .
\label{har}
\eea
For the curvature, keeping the linear terms,  
\bea
R \, = \, \Box \, h - \partial_\mu h^\mu \, = \, 
- \, \dfrac{1}{2} \, \Box \, \chi - \partial_\mu \chi^\mu
\quad , \quad  \text{where} \quad h^\mu \equiv \partial_\nu h^{\mu \nu} .
\eea
In this $\, \Box$ is  the flat metric d' Alembertian  operator.  
In the harmonic gauge
\bea
R^{\mu \nu} - \dfrac{g^{\mu \nu} }{2} \, R \, = \, \dfrac{1}{2} \, \, \Box \, \chi^{\mu \nu} ,
\eea
and also ( in harmonic gauge )
\bea
&&( \,  g^{\mu \nu}  \, \Box \, - \,    \nabla^\mu   \,  \nabla^\nu  \,   ) \,  R \, \quad 
\, = \, - \,  \dfrac{1}{2} \, ( \,  n^{\mu \nu}  \, \Box \, - \,    \nabla^\mu   \,  \nabla^\nu  \,   ) \, \Box \, \, \chi
\nonumber \\
&&( \,  g^{\mu \nu}  \, \Box \, - \,    \nabla^\mu   \,  \nabla^\nu  \,   ) \, \Box \,  R \,
\, = \, - \,  \dfrac{1}{2} \, ( \,  n^{\mu \nu}  \, \Box \, - \,    \nabla^\mu   \,  \nabla^\nu  \,   ) \, \Box^{\, 2}  \, \chi .
\eea

\section{Mass spectrum and allowed range of the parameters}  \label{app3}

The mass spectrum, arising from the mixing of the graviton with the real part of $\, F_q$,  includes two massless states, the graviton in  two helicity states, and two massive Klein-Gordon fields. 
 The masses squared $\,  m_1^2 \, , \, m_2^2$ are the eigenvalues of the mass matrix appearing in (\ref{diagsys}) and hence  they are given by
 \bea
  m_{1,2}^2 \, = \, \dfrac{ \,   ( \rho_2 + \sigma_1) \pm \sqrt{{ ( \rho_2 - \sigma_1)^2 } + 4 \, \rho_1 \, \sigma_2 } \, }{2}
  \quad \text{with} \quad m_1^2 > m_2^2 .
 \label{masses}
 \eea
 These should be real and nontachyonic, which  restricts the available range of the parameters involved.  
However  there are ranges, as we shall discuss later,  where these conditions can be simultaneously satisfied. 
Their  masses squared    in (\ref{masses})   can be more conveniently expressed in the following manner,
 \bea
  m_{1,2}^2 \, = \, \dfrac{ \,   B \pm \left( { B^2 - 4 \, {\alpha} \, ( 4 \, \beta \, \lambda - 1   ) \, f_0 )  } \right)^{1/2}\, }{2 \alpha ( 4 \, \beta \, \lambda - 1   ) }
  \quad \text{with} \quad m_1^2 > m_2^2 .
 \label{masses2}
 \eea
 In this, $\, B$ is given by
 \bea
 B \, = \, \lambda \, (f_0^\prime )^2  - \,  f_0^\prime + 4 \, \alpha \, \lambda \, f_0 \, + \, \beta .
 \label{BBB}
 \eea
 To arrive at (\ref{masses2}) we have used the fact that the following relations hold, due to the relation (\ref{frrr}),
 \bea
 F(0) = 0 \quad , \quad F^\prime(0) = - \, \dfrac{f(0)}{3} \quad , \quad 
 F^{\prime \prime}(0) = - \, \dfrac{f^\prime(0)}{9} \, + \, \dfrac{\beta}{9} .
 \label{zeror}
 \eea
 
 To avoid tachyonic solutions, the  sum and the product of the eigenvalues  (\ref{masses2}) should be positive. These  lead to the conditions
 \bea
 ( 4 \, \beta \, \lambda - 1   ) \, B \, > \, 0    \quad , \quad ( 4 \, \beta \, \lambda - 1   ) \, f_0 \, > \, 0 .
 \eea
 On the other hand, the reality of the masses imposes  
 \bea
 B^{\, 2} \, \, > \,  \, 4 \, \alpha \,  ( 4 \, \beta \, \lambda - 1   )  \, f_0 .
 \eea
These conditions have to be simultaneously  satisfied. 
To ensure that a range of the parameters exists where this holds, consider  the following range
\bea
f_0 > 0  \quad , \,  \quad   f_0^\prime \,  >  \dfrac{1}{  \, \lambda} 
\quad , \quad 4 \, \beta \, \lambda > 1
\quad   ,  \quad \beta \quad  \text{sufficiently large}  , 
\label{cond3}
\eea
Certainly, one can find other ranges, as well, but it suffices to consider this range of the parameters. 
Note that positivity of $f_0$ is also demanded in order to have, in the dual $\, F(R)$-theory, a linear in the curvature term having  the correct sign  and the condition $ 4 \, \beta \, \lambda > 1 $ is necessary in order to avoid ghosts in the standard $N=1$ supergravity, as we have discussed in the main text. 

As for the masses of the $\, \Psi, G$ system,  they  are expressed in terms of
$\, \delta_{11} , \delta_{22}  $ and $\, \delta_{12}, \delta_{21}  $  appearing in the mass matrix of Eq.  
(\ref{diagsys2}), in the following manner
 \bea
  m_{\pm}^2 \, = \, \dfrac{ \,   ( \delta_{11} + \delta_{22}) \pm \sqrt{{ ( \delta_{11} + \delta_{22}) ^2 } -
  4 \, ( \delta_{11} \, \delta_{22}  - \delta_{12} \, \delta_{21}  ) }   \, }{2} .
 \label{masses3}
 \eea
 It is not difficult to see that,  
the pertinent  combination  defining the masses  in  (\ref{masses3}) above , are given by 
\bea
 \delta_{11} + \delta_{22}  \, &=& \,  \, \dfrac{B}{  \, \alpha \, D} 
 \label{con41}
  \\
  \; \delta_{11} \, \delta_{22}  - \delta_{12} \, \delta_{21} \, &=& \, \dfrac{f_0}{ \, \alpha \, D}  ,
 \label{con44}
 \eea
 where the parameters $\, B $ is exactly  (\ref{BBB}) and $D$ is given by, 
 \bea
 D \, = \, 4 \, \beta \, \lambda \, - \, 1 .
 \label{DDD}
 \eea
 In terms of these, the masses squared are analytically given by
 \bea
 m_{\pm}^2 \, = \, \dfrac{  B \,  \pm   \, \left( \,  B^2 - 4\, \alpha \, f_0 \, D     \,  \right)^{1/2} }{2 \, \alpha \,  D } \; ,
 \label{masses33}
 \eea
 which coincide with the masses given in (\ref{masses2}).

 The masses that arise from the system of the fields $\, A , \rho$ is found by studying Eqs. (\ref{aro3}),  which is expressed in terms of the parameters 
 \bea
 c_1 \, &=& \, - \, \dfrac{1}{4 \, \alpha \lambda} \quad \quad  \quad  , \quad \quad  c_2 \, = \,  \, \dfrac{1}{4 \, \alpha \lambda} \, - \, 
\dfrac{f_0^\prime}{2 \, \alpha}
\nonumber \\
g \, \; &=& \, \dfrac{ 4 \lambda \, \beta - 1   }{  1 \, - \, 2 \lambda {f_0^\prime} } 
\quad \quad , \quad  \quad \mu \, = \, - \, \dfrac{ 4 \, \lambda \,  f_0  }{ 1 \, - \, 2 \lambda {f_0^\prime} } 
\label{stath55}
 \eea 
 The mass  matrix $\,  \pmb{ {\cal{M}}^2} $ is  determined by the matrix elements given below
 \bea
 m_{11} \, = \, \dfrac{ c_2 - \mu}{ g  } \quad , \quad  m_{12}  = \dfrac{c_1 }{ g}  \quad , \quad  m_{21} = - \,  c_2 \quad , \quad 
   m_{22} \, = \, - \, c_1 .
 \label{elem}
 \eea
  The secular equation determining the eigenvalues of $\,  \pmb{ {\cal{M}}^2} $ is given by
  \bea
  \xi^2 - S \, \xi + P \, = \, 0 ,
  \label{secul}
  \eea
  with $S , P$, the sum and the product of the eigenvalues. These are found to be 
\bea
 S \, = \, \dfrac{ B }{ \alpha  \, D  } \, 
 \; , \;
 P \, = \, \dfrac{ f_0 }{ \alpha \, D  }  . 
 \label{sumpro}
\eea
These coincide with (\ref{con41}) and (\ref{con44}). That is the eigenvalues in this case  have the same sum and product as in the previously considered  system.  Therefore the masses are identical to  (\ref{masses33}), and hence  (\ref{masses2}).

 \section{The scalar potential } \label{appot}
 In the presence of the stabilizer (\ref{kallosh}) the complete form of the scalar  potential is given by 
 \bea
 V \, = \, \dfrac{ 9 }{ 2 \lambda \, \epsilon \, \Omega^2  } \, P .
 \label{potall}
 \eea
 Where the function $P$ is field depended, given by, 
\bea
P \, = \, & & | C - 2 \lambda \, ( T +f(C) ) |^{\,2} 
 \nonumber \\ 
 & &
 + \, \epsilon\, \Big(  \, | C |^{\,2} + \dfrac{\lambda}{\alpha} \,   | \, \Phi f^\prime(C) + Q  \, | ^2  + 2 \lambda \, ( \, C Q \overline{\Phi} + h.c. \, )  \,   +  2 \lambda  \, ( \, T  + 2  f(C)  -  C f^\prime(C) + h.c. \, ) \, \, | \Phi |^2 \, \Big)
\nonumber \\ 
& &
+ \, 2 \lambda \, \zeta  \,
 | \Phi |^4 \, ( \, - 4 \lambda \, T  +  2 \,C - 4 \lambda  \, f(C)   + h.c. \, ) \, + \,
  \, 2 \lambda \, \zeta  \, ( \, 4 \beta \lambda - 1 ) \,  | \Phi |^6 \, .
  \label{pfun}
\eea
The expression $\epsilon$, appearing in the equations above, is given by
\bea
\epsilon \, = \, 4 \beta \lambda - 1 - 8 \zeta \, \lambda \, | \Phi |^2 .
\label{eee}
\eea
In the limit $\zeta =0$ the last two terms of $P$  vanish while $\epsilon$ becomes $ 4 \beta \lambda - 1 $. Then we recover the potential given by Eq. (\ref{potit}).  
Recall that in supergravity theories we need have $det \, {\cal{K}}_{\bar{J} I} > 0 $, as well as $\Omega < 0$,  and this leads to 
$\epsilon > 0$. Actually   the latter  is proportional to  $\, - \,  \Omega^5 \, det \, {\cal{K}}_{\bar{J} I} / 162 \alpha $, and thus positive due to the fact that $\Omega < 0$. 

The minima of the potential are found by solving $\, \partial_i V =0$ which yields
\bea
- \dfrac{1}{( \epsilon \Omega^2 )^2} \, (  \partial_i \epsilon \Omega^2) \, P \, + \, \dfrac{1}{ \epsilon \Omega^2 } \, \partial_i \, P \, = \, 0 .
\label{extr}
\eea
It is easy to verify that at the field values
\bea
ReT = - f_0, \; ImT = 0, \, C = \Phi = Q = 0 
\label{extrem2}
\eea
(\ref{extr}) is  satisfied. In particular  at this point both 
\bea
P = 0 \quad , \quad \partial_i \, P \, = \, 0 .
\label{pderp}
\eea
The point (\ref{extrem2}) is indeed a minimum  as we have shown in the main text. In fact  the masses squared of the associated scalar fields are all  positive, in some range of the parameters involved [ see \ref{quadpot} and discussion following it ]. 

In the following we shall prove the there are values of $\zeta$ for which the potential is positive semidefinite for any values of the fields involved. To that purpose, as is evident from (\ref{potall}), only positivity of the function $P$, given by (\ref{pfun}),  is required.  
 With $ P \geq 0$ it is  guaranteed that (\ref{extrem2})  is the absolute minimum.
 
The proving  task  is facilitated if one uses the following field combinations 
\bea
Q^\prime = Q + \Phi f^\prime(C) \quad , \quad U = C - 2 \lambda( T + f(C) )  .
\eea
The advantage of using these combination is that the function $P$ receives the form
\bea
P \, = \, | U |^2 + {B} \, ( U + \overline{U})  + \Gamma ,
\eea
which can be   cast as 
\bea
P \, = \, \left( Re U + {B} \right)^2  + \left( Im U \right)^2 + \left( \Gamma - {B^2} \right) .
\label{pex}
\eea
The functions $ B, \Gamma $ are explicitly  given by
\bea
B &=& -  \,( 4 \beta \lambda - 1 - 12 \lambda \zeta | \Phi |^2 ) \,  | \Phi |^2
\\ \nonumber
\Gamma &=&
\epsilon\, \Big(  \, | C |^{\,2} + \dfrac{\lambda}{\alpha} \,   | \, Q^\prime  \, | ^2  + 2 \lambda \, ( \, C Q^\prime \overline{\Phi} + H.c. \, )  \,   +    \, ( \, F(C) + H.c. \, ) \, \, | \Phi |^2 \, \Big) 
\\ \nonumber
&& + \, 2 \lambda \, \zeta  \, ( \, 4 \beta \lambda - 1 ) \,  | \Phi |^6 \, .
\eea
In the definition of $\Gamma$ the function $F(C)$ is the analytic function
\bea
F(C) \, = \, C + 2 \lambda ( \, f(C) - 2 \, C\, f^\prime(C) ) .
\label{FFff}
\eea

In order to proceed further we express the magnitude squared  of the field $\Phi$ in terms of $\epsilon, \zeta$,
\bea
\rho \equiv |\Phi |^2 \, = \, \dfrac{4 \beta \lambda - 1  - {\epsilon}}{8 \lambda \zeta} ,
\eea
and replace it into Eq. (\ref{pex}). That done,  $ P $ receives the form
\bea
P &=& 
 \left( Re U + {B} \right)^2  + \left( Im U \right)^2 
+ \,
\dfrac{ \epsilon  \, \lambda}{\alpha} \, ( \,  |Q^\prime | + 2 \, \alpha \, cos( \theta - \omega ) \, | C | \, \sqrt{\rho} \,  )^2
\, + \, 
 \nonumber \\
& & \; \; \; \; + \epsilon \, \Big( \; ( 1 - 4 \alpha \, \lambda  \, cos^2( \theta - \omega ) \, \rho ) | C |^2 \, + \, 2 Re  F(C)  \, \rho
\, + \, \rho \, \overline\delta \; \Big) .
\label{pmin2}
\eea

The angles $\theta, \omega$ in Eq. (\ref{pmin2}) are not actually needed, for the discussion that follows, but for reasons of completeness we state that they are the arguments of $\Phi$ and $ Q^\prime C$ complex fields respectively, i.e.,
\bea
\Phi = | \Phi | \,  e^{\, i \theta } \quad , \quad  Q^\prime C = | Q^\prime C | \, e^{\, i \omega } .
\eea
The   $\overline\delta$ in the  last term of (\ref{pmin2}) is given by
\bea
\overline\delta \, = \,   \dfrac{ 9 \epsilon^2 - 14 ( 4 \beta \lambda - 1  ) \epsilon + 5 \,( 4 \beta \lambda - 1 )^2 }{ 32 \lambda \zeta } 
\eea
which  has a minimum, as function of $\epsilon$,  given by 
\bea
\overline\delta_{min} \, = \,  - \dfrac{ \,   ( 4 \beta \lambda - 1 )^2}{72 \, \lambda \, \zeta} \, .
\eea

In   Eq. (\ref{pmin2}), the first three terms are positive, hence  a lower bound can be established given by 
\bea
P  \geq   
 \epsilon \, \Big( \; ( 1 - 4 \alpha \, \lambda \, cos^2( \theta - \omega )  \, \rho ) | C |^2 \, + \, 2 Re  F(C)  \, \rho
\, + \, \rho \, \overline\delta \; \Big) \, \equiv \epsilon \, b_P .
\label{pmin22}
\eea
The important thing is that, 
if we manage to make  the function $ b_P$,  appearing on the right-hand side  of the equation  above,    positive, for any values of the fields involved,  then the potential will be positive too. 
When $\rho=0$, corresponding to $\Phi=0$,  the potential  is  positive semidefinite since  the bound above becomes 
$ P \geq \epsilon \,  | C |^2 \geq 0$ .  When $\rho \neq 0$,  we shall show that this is indeed the case  for sufficiently large values of the parameter $\zeta$,
provided $F(C)$ is at most quadratic in the field $C$.  Hence in either case we would have  $P \geq  \ 0$ and thus the potential would be  positive semidefinite. The proof goes as follows.

For any analytic function $F(C)$, and hence $f(C)$, which is at most quadratic in $C$, we write
\bea
F(C) = a_0 + a_1 C + a_2 C^2 \equiv 2 \lambda f_0 +  | a_1 | | C | \, e^{i \theta_1} +  | a_2 | | C |^2 \, e^{i \theta_2} .
\label{trion}
\eea
In this we have used the fact that the constant term $a_0$ is  $ 2 \lambda f(0) = 2 \lambda f_0$, using Eq. ( \ref{FFff}), and the fact  that $f_0$ has been taken real. Using this the function $b_P$, on the right of Eq. (\ref{pmin22}),  can be written as 
\bea
b_P \, = \, 
 \; ( 1 - 4 \alpha \, \lambda \, cos^2( \theta - \omega )  \, \rho + 2 | a_2| \,  cos \theta_2 \, \rho \,) | C |^2 \, + \, 2 | a_1 | \, cos \theta_1  \, \rho \, | C |
\, + \, \rho \, ( 4 \lambda f_0  \, + \,  \overline\delta ) ,
\eea
which is bounded from below as follows,
\bea
b_P \geq 
( 1 - ( 4 \alpha \, \lambda \,   + 2 | a_2| \, ) \, \rho \,) | C |^2 \, - \, 2 | a_1 |   \, \rho \, | C |
\, + \, \rho \, ( 4 \lambda f_0  \, + \,  \overline\delta_{min} ) \, \equiv \, b_P^{min} .
\label{pmin333}
\eea
The bound $ b_P^{min} $   is a polynomial of second degree in $|C|$.  For  sufficiently small values of $\rho$,  corresponding to large values of $\zeta$,  the coefficient of $ | C |^2$ can become  positive. On the other hand, its discriminant is given by
\bea
{\cal{D  }} \, = \, 4  \, ( \, | a_1|^2 \,  + \,  ( 4 \alpha \, \lambda \,  + \, 2 |a_2| \, )   \, ( 4 \lambda f_0 + \overline\delta_{min} )  ) \, \rho^2 \, - \, 4 ( 4 \lambda f_0 + \overline\delta_{min} )  \, \rho .
\label{dis}
\eea
The first term of it is proportional to $\rho^2$ and the second  proportional to $\rho$. Hence the latter dominates for 
sufficiently small $\rho$, or same, for sufficiently large values of $\zeta$.  Therefore for such values of $\zeta$ the sign of 
${\cal{D  }} $  is dictated by the sign of the last term of (\ref{dis}). This is negative when  
$ 4 \lambda f_0 + \overline\delta_{min}  > 0  $ resulting to, 
\bea
\zeta >  \dfrac{ \, ( 4 \beta \lambda - 1 )^2}{288 \, f_0 \, \lambda^2  } .
\label{rang}
\eea
Therefore there are always adequately large values of $\zeta$, for which   ${\cal{D  }} $ is negative and the coefficient of $ | C |^2  $ is positive. In this regime  $b_P^{min}$ in Eq. (\ref{pmin333}), and hence $b_P$, is always positive and the scalar  potential is positive semidefinite. 

In order to better quantify the bounds imposed on $\zeta$, one can see that both conditions, that is  positivity of the  coefficients of 
the $|C|^2$ term, and  ${\cal{D  }} < 0 $ yields,  
\bea
\dfrac{1}{\rho} \geq 4 \alpha \, \lambda \,  + \, 2 |a_2|  + \dfrac{ | a_1|^2}{ 4 \lambda f_0 + \overline\delta_{min}  } ,
\label{bound1}
\eea
when $ 4 \lambda f_0 + \overline\delta_{min}  > 0  $.  Inequality (\ref{bound1})  is always  satisfied when $\zeta$ lies in the range
\bea
\dfrac{8 \lambda  \zeta }{ 4 \beta \lambda - 1 } \geq 4 \alpha \, \lambda \,  + \, 2 |a_2|  + 
\dfrac{ | a_1|^2}{ 4 \lambda f_0 + \overline\delta_{min}  } .
\label{bound2}
\eea
This leads to the condition that a quadratic  polynomial, in the parameter  $\zeta$,   is positive.  In it, the coefficient of the  $\zeta^2$ is positive and we have verified that it has  two real roots, the largest of these being $r_+$.  
This involves the constants  $a_1, a_2  $ which define the function $F(C)$ in (\ref{trion}).
Therefore,  (\ref{bound2})  holds for $\zeta > r_+$. 

Combining  the two bounds,  (\ref{rang}) and (\ref{bound2}), we get
\bea
\zeta \geq  max \Bigl\{ \,  r_+ \, ,  \dfrac{  ( 4 \beta \lambda - 1  )^2  }{  288 f_0 \lambda^2}   \Bigr\}
\eea
  
  We conclude by saying  that, for any function $f(C)$, and hence $F(C)$,   at most  quadratic in $C$, the stabilization constant 
  $\zeta$ can be taken sufficiently large so that the scalar potential is positive semidefinite. In this case the minimum given in  (\ref{extrem2}), for which the potential vanishes,  is the absolute minimum.


\end{document}